\documentclass{article}

\usepackage{arxiv}

\usepackage[utf8]{inputenc} 
\usepackage[T1]{fontenc}    
\usepackage[hidelinks]{hyperref}       
\usepackage{url}            
\usepackage{booktabs}       
\usepackage{amsfonts}       
\usepackage{nicefrac}       
\usepackage{microtype}      
\usepackage{lipsum}         
\usepackage{graphicx}
\usepackage{natbib}
\usepackage{doi}
\usepackage{dsfont}
\usepackage{amsmath}
\usepackage{amsthm}
\usepackage{amssymb}
\usepackage[ruled]{algorithm2e}
\usepackage{cleveref}       

\makeatletter
\theoremstyle{plain}
\newtheorem{thm}{\protect\theoremname}
\theoremstyle{plain}
\newtheorem{prop}[thm]{\protect\propositionname}
\theoremstyle{plain}
\newtheorem{fact}[thm]{\protect\factname}
\theoremstyle{plain}

\theoremstyle{plain}
\newtheorem{cor}[thm]{\protect\corollaryname}
\theoremstyle{definition}
\newtheorem{defn}[thm]{\protect\definitionname}
\theoremstyle{definition}
\newtheorem{ex}[thm]{\protect\examplename}
\theoremstyle{definition}
\newtheorem{open_q}[thm]{\protect\openquestionname}
\theoremstyle{plain}
\newtheorem{lem}[thm]
{\protect\lemmaname}

\makeatother

\usepackage{babel}
\providecommand{\conjecturename}{Conjecture}
\providecommand{\corollaryname}{Corollary}
\providecommand{\definitionname}{Definition}
\providecommand{\examplename}{Example}
\providecommand{\factname}{Fact}
\providecommand{\propositionname}{Proposition}
\providecommand{\theoremname}{Theorem}
\providecommand{\definitionname}{Definition}
\providecommand{\openquestionname}{Open Question}
\providecommand{\lemmaname}{Lemma}

\title{The occlusion process: improving sampler performance with parallel computation and variational approximation}

\newif\ifuniqueAffiliation
\uniqueAffiliationtrue

\ifuniqueAffiliation 
\author{\href{https://maxhhird.github.io/}{Max Hird}\\
	Department of Statistical Science\\
	University College London\\
	1–19 Torrington Place, London WC1E 7HB \\
	\texttt{max.hird.19@ucl.ac.uk} \\
	\And
	\href{https://dms.umontreal.ca/en/departmental-directory/professors/portrait/maire}{Florian Maire} \\
	D\'{e}partement de math\'{e}matiques et de statistique\\
	Universit\'{e} de Montr\'{e}al\\
	Andr\'{e}-Aisenstadt Local 4253 \\
    \texttt{florian.maire@umontreal.ca} \\
}
\else
\usepackage{authblk}

\newbox{\orcid}\sbox{\orcid}{\includegraphics[scale=0.06]{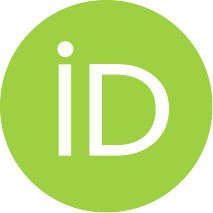}} 
\author[1]{%
	\href{https://orcid.org/0000-0000-0000-0000}{\usebox{\orcid}\hspace{1mm}David S.~Hippocampus\thanks{\texttt{hippo@cs.cranberry-lemon.edu}}}%
}
\author[1,2]{%
	\href{https://orcid.org/0000-0000-0000-0000}{\usebox{\orcid}\hspace{1mm}Elias D.~Striatum\thanks{\texttt{stariate@ee.mount-sheikh.edu}}}%
}
\affil[1]{Department of Computer Science, Cranberry-Lemon University, Pittsburgh, PA 15213}
\affil[2]{Department of Electrical Engineering, Mount-Sheikh University, Santa Narimana, Levand}
\fi

\renewcommand{\headeright}{}
\renewcommand{\undertitle}{}
\renewcommand{\shorttitle}{The occlusion process}

\hypersetup{
pdftitle={A template for the arxiv style},
pdfsubject={},
pdfauthor={Max Hird, Florian Maire},
pdfkeywords={Markov chain Monte Carlo, Variational Inference, Parallel Computation},
}

\begin{document}
\maketitle

\begin{abstract}
	Autocorrelations in MCMC chains increase the variance of the estimators they produce. We propose the occlusion process to mitigate this problem. It is a process that sits upon an existing MCMC sampler, and occasionally replaces its samples with ones that are decorrelated from the chain. We show that this process inherits many desirable properties from the underlying MCMC sampler, such as a Law of Large Numbers, convergence in a normed function space, and geometric ergodicity, to name a few. We show how to simulate the occlusion process at no additional time-complexity to the underlying MCMC chain. This requires a threaded computer, and a variational approximation to the target distribution. We demonstrate empirically the occlusion process' decorrelation and variance reduction capabilities on two target distributions. The first is a bimodal Gaussian mixture model in 1d and 100d. The second is the Ising model on an arbitrary graph, for which we propose a novel variational distribution.
\end{abstract}


\tableofcontents

\section{Introduction}\label{section:intro}

One of the principal motivations to use a Markov chain Monte Carlo
(MCMC) method is to estimate expectations with respect to a given
distribution $P$. This is done by averaging functionals over the
chains which the MCMC method produces. It is well known that autocorrelations
in the chain result in increased variance in the ensuing estimator. Our proposed algorithm, the \emph{occlusion
process}, combines MCMC with a variational approximation to the target with the
aim of achieving consistent estimators with reduced variance.

Specifically, the occlusion process takes as input a partition of the state space
into disjoint regions and is constructed upon an existing
MCMC sampler. It monitors the Markov chain produced by the
sampler along with the region it is in and, where possible, it produces
a sample from $P$ restricted to the region. Upon this event it will
use the sample instead of the state from the Markov chain
in the estimator, hence \emph{occluding} the Markov chain state from view. 

Compare the variance of a functional averaged over the run of a positive Markov chain in a given region, and the variance of that functional averaged over independent draws from $P$ restricted to that region. It is clear that the latter will be smaller than the former. This is the way in which the occlusion process aims to reduce variance.

The process uses a variational distribution, and is
designed to exploit parallel computation to boost its performance. Given
a variational distribution and a threaded computer, the process is
straightforward to implement, and its running time is the same as
the MCMC sampler it is constructed on by design.

\begin{figure}[h]
    \centering
    \includegraphics[scale = 2.5]{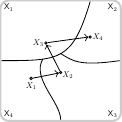}\quad\includegraphics[scale = 2.5]{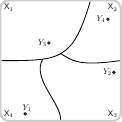}\quad\includegraphics[scale = 2.5]{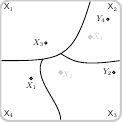}
    \caption{Three versions of the state space $\mathsf{X}$; the leftmost with the Markov chain $\{X_t\}$ and the middle with the samples $\{Y_t\}$ taken from the target restricted to the regions that the Markov chain visits. We assume that we were only able to successfully sample $Y_2$ and $Y_4$, therefore the rightmost picture shows the samples we will use for the occlusion estimator: $X_2$ and $X_4$ have been \emph{occluded} by $Y_2$ and $Y_4$.}
    \label{fig:state_spaces}
\end{figure}

\begin{figure}[h]
    \centering
    \includegraphics[scale = 1.25]{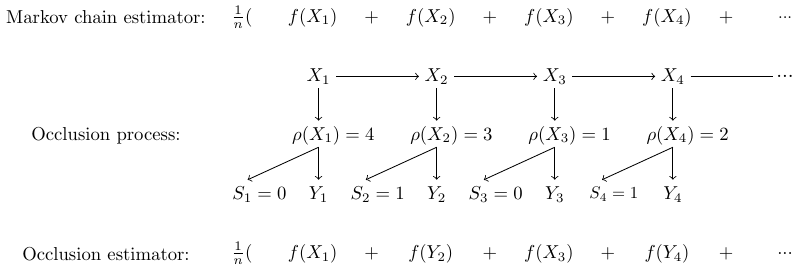}
    \caption{The top line is the estimator constructed using states of the Markov chain $\{X_t\}$. The middle picture is a DAG representing the occlusion process: $\{X_t\}$ is the Markov chain, ${\rho(X_t)}$ denotes the regions visited by the Markov chain, $\{Y_t\}$ are the samples from the target restricted to those regions, and $\{S_t\}$ indicate which of those samples we were able to successfully produce. The bottom line is the occlusion estimator made up of the samples from the Markov chain, and the successfully produced $Y_t$'s.}
    \label{fig:DAG_diagram}
\end{figure}

Figures \ref{fig:state_spaces} and \ref{fig:DAG_diagram} offer pictorial representations of the process: in Figure \ref{fig:state_spaces} we see three versions of the state space $\mathsf{X}$ partitioned into $\{\mathsf{X}_1, \mathsf{X}_2, \mathsf{X}_3, \mathsf{X}_4\}$. The leftmost picture shows the beginning of an MCMC chain visiting regions 4, 3, 1, and 2 in that order. The occlusion process monitors the Markov chain, and attempts to produce samples $Y_1, Y_2, Y_3,$ and $Y_4$ from the target distribution restricted to the regions the Markov chain has just visited. These samples are shown in the middle picture. Since we are in the context of MCMC, the target distribution will be such that we will not be able to successfully produce \emph{all} of these samples within some bounded time horizon. Let's say we were only able to produce $Y_2$ and $Y_4$ in a reasonable amount of time. The rightmost picture then shows which samples will be used in the estimator: $X_1, Y_2, X_3,$ and $Y_4$. Samples $Y_2$ and $Y_4$ have therefore \emph{occluded} samples $X_2$ and $X_4$.

Figure \ref{fig:DAG_diagram} has, from top to bottom, the estimator which uses the Markov chain samples, a DAG demonstrating the occlusion process, and the estimator after the occlusions. It relates to the example process depicted in Figure \ref{fig:state_spaces}. The figure contains two additional pieces of information to Figure \ref{fig:state_spaces}: we denote by $\rho(X_t)$ the region that $X_t$ is in, and $S_t$ is an indicator which indicates the successful production of a sample $Y_t$ from the target, restricted to $\mathsf{X}_{\rho(X_t)}$. It demonstrates the following properties of the occlusion process: that $\{S_{t}\}\bot\{X_{t}\}\left|\{\rho(X_{t})\}\right.$, $\{S_{t}\}\bot\{Y_{t}\}\left|\{\rho(X_{t})\}\right.$, and $\{Y_{t}\}\bot\{X_{t}\}\left|\{\rho(X_{t})\}\right.$. In addition we have that $Y_t \bot Y_{t'}\left|\{\rho(X_{t})\}\right.$ and $S_t \bot S_{t'}\left|\{\rho(X_{t})\}\right.$ for all $t \neq t'$. The figure shows that the occlusion process can be viewed as a hidden Markov model. Note also that at no point do we assume that the chain is regenerative, we simply need it to be invariant to the target distribution $P$.

Variance reduction methods constitute a historied subgenre of computational statistics. They take existing algorithms and modify them
to produce estimators with the same expectation but reduced variance.
The change in computational work between the algorithm and its reduced
variance modification will be sufficiently small such that an overall
increase in efficiency is achieved. The variance reduction method
most similar to the method we devise here is stratified sampling with
proportional allocation \cite[Chs 5 - 5A]{cochran:1977}. The occlusion process is similar
to stratified sampling because the aforementioned partition acts as
a stratification, and the proportional allocation comes from the fact
that the number of samples used in the estimator from $P$ restricted to a
particular region will be proportional to the length of time the Markov
chain spends in that region. What separates our method from stratified
sampling is that the conditions for stratified sampling are far more
restrictive. Namely, stratified sampling assumes we can sample from
$P$ restricted to every region and it assumes that we know the mass
of every region under $P$. It is known that stratified sampling with
proportional allocation produces lower variance estimators than plain
Monte Carlo (MC). One might expect to extend this result to our less
restrictive context, although interestingly, we find a counterexample.

Variational inference (VI) \citep{blei:2017} consists in finding
a distribution $Q$ which is close to $P$ in some appropriate sense,
but that has desirable properties. These properties might include,
for instance, the ease with which we sample from it. One of the limitations
of VI is that, depending on the optimisation procedure to search for
$Q$ and the family of distributions we restrict the search to, samples
from $Q$ will not be samples from $P$. Therefore previous authors, e.g.
\citep{parno:2018, brofos:2022}, have sought to
incorporate $Q$ into (asymptotically) exact methods such as MCMC.
This is also the case with a particular implementation of our method, which uses $Q$ as the proposal
distribution in rejection samplers whose samples are used to occlude
those of a Markov chain.

Parallel computing allows users to perform more computational operations
in a fixed length of time. MCMC is a prima facie serial operation,
and hence exploiting parallel threads is an attractive prospect since
we are able to do additional computations alongside the Markov chain,
which may need to be run for a long time to reach equilibrium. These
additional computations may take the form of other Markov chains \citep{surjanovic:2023, hoffman:2021}. In our case we devote the additional computing capacity
to the rejection samplers mentioned above.

It should be noted that the occlusion process shares some similarities with the Kick-Kac samplers conceived in \citep{douc:2023}. These samplers partition $\mathsf{X}$ into two measurable regions $\{\mathsf{X}_1, \mathsf{X}_2\}$, and they attempt to take independent samples from $P$ restricted to $\mathsf{X}_1$. Absent this ability, they instead use a Markov chain which is invariant to $P$ restricted to $\mathsf{X}_1$. Kac's theorem \citep{kac:1947} dictates that if we average a functional over a $P$-invariant Markov chain, beginning at one of these independent samples and ending when it re-enters $\mathsf{X}_1$, we get an unbiased estimator of the expectation of the functional with respect to $P$.

Whilst the setup is very similar to that of the occlusion process, we note the following differences. Firstly the Kick-Kac samplers use two regions, each with a different purpose. The occlusion process uses any number of regions, and they are all treated in the same way. Secondly, to produce samples from $P$ restricted to $\mathsf{X}_1$, the Kick-Kac samplers must either wait for an independent sample, or use a Markov chain which is invariant to it. Therefore the process either has to wait for what is possibly a long time, or introduce autocorrelations. When the occlusion process achieves a sample $Y_t$ from $P$ restricted to a region, it is guaranteed to be independent from all other random variables given $\rho(X_t)$ (see Figure \ref{fig:DAG_diagram}). The process itself is not fully dependent on these samples, and so can continue regardless of the probability of their production. Thirdly the Kick-Kac samplers rely on regenerations to build their estimators and so they have a random time complexity, whereas the occlusion process does not rely on such conditions and the time complexity is specified before the start of the algorithm.

\subsection{Notation}\label{subsection:notation}

We denote by $\mathsf{X}$ the state space of the target measure $P$ and $\mathcal{X}$ the Borel $\sigma$-algebra on $\mathsf{X}$. For all $p \in [1, \infty)$ we define the space $L^p(P)$ of functions $f : \mathsf{X} \to \mathbb{R}$ as $L^p(P) := \{f: \mathbb{E}_P[|f(X)|^p]^\frac{1}{p} \leq \infty\}$. We endow $L^2(P)$ with the inner product $\langle f, g \rangle_P := \mathbb{E}_P[f(X)g(X)]$ for all $f,g\in L^2(P)$. The associated norm is then $\|f\|_P := \langle f, f \rangle_P^\frac{1}{2}$. For a given measure $\mu$ and a function $f$ we use the notation $\mu (f) := \mathbb{E}_\mu[f(X)]$ where the expectation exists. Say $K:\mathsf{X} \times \mathcal{X} \to [0, \infty)$ is a Markov kernel. It is defined to act on a function $f:\mathsf{X} \to \mathbb{R}$ with $Kf(x) := \int_\mathsf{X} f(x')K(x \to dx')$ for all $x \in \mathsf{X}$ and a probability measure $\mu:\mathcal{X} \to [0, 1]$ with $\mu K(A) := \int_\mathsf{X} \mu(dx)K(x \to A)$ for all $A \in \mathcal{X}$.

For a given $R \in \mathbb{N}\backslash\{0\}$ we define $[R] := \{1,2,...,R\}$. Say $\mathsf{X}$ is partitioned into $\{\mathsf{X}_i:i\in [R]\}$ such that $P(\mathsf{X}_i)>0$ for all $i\in [R]$. We define $P_i$ to be $P$ restricted to $\mathsf{X}_i$ for all $i \in [R]$. Given a function $f\in L^2(P)$ we define $\mu_i$ and $\sigma_i^2$ to be $P_i(f)$ and $P_i((f - P(f))^2)$ respectively. Given a size $R$ partition, we define the function
\begin{equation*}
    \rho(x) := \sum_{i = 1}^R i \mathds{1}\{x \in \mathsf{X}_i\}
\end{equation*}
that simply outputs which part $x \in \mathsf{X}$ is in. 

Given three random variables $X$, $Y$, and $Z$ we denote by $X \perp Y | Z$ the independence of $X$ and $Y$ given $Z$. We denote the non-negative reals by $\mathbb{R}^+$. For a set $A$ we denote by $\mathcal{P}(A)$ its power set. For a measurable set $A \in \mathcal{X}$ we define $-A := \{-a:a\in A\}$.

\subsection{Paper Structure}\label{subsection:paper_structure}

In all this paper serves as an introduction to, and a proof of concept of, the occlusion process.

In Section \ref{section:the_occlusion_process} we introduce the occlusion process, define it mathematically, and state its properties. In Section \ref{subsection:stratified_sampling} we introduce stratified sampling, as it is closely related to the occlusion process. In Section \ref{subsubsection:the_fully_occluded_estimator} we introduce an idealised version of the occlusion process whereby we assume the ability to sample from each $P_i$ at minimal computational cost. We define the ensuing estimator and examine its variance properties from Proposition \ref{prop:ideal_estimator} onwards. This bridges the gap between stratified sampling and Section \ref{subsubsection:the_occluded_estimator} where we define the occlusion process in equation (\ref{eqn:K_occ_definition}) and the corresponding occluded estimator in equation (\ref{eqn:mu_hat_occ}). We define its equilibrium distribution in equation (\ref{eqn:P_occ_definition}) and state the variance of the occluded estimator in Proposition \ref{prop:variance_mu_occ}.

The occlusion process is built upon an underlying $P$-invariant Markov chain, such as one produced by an MCMC algorithm. In Section \ref{section:inherited_theoretical_properties_of_the_occlusion_process} we examine which properties the occlusion process will inherit from the underlying Markov chain. In Section \ref{subsubsection:basic_properties} we see that it inherits reversibility, amongst other basic properties. In Section \ref{subsection:law_of_large_numbers} Theorem \ref{thm:LLN_inheritance} we see that it inherits a Law of Large Numbers. In Section \ref{subsection:convergence_in_a_given_norm} Theorem \ref{thm:inherited_normed_function_convergence} we see that it inherits convergence in a normed function space, and in Section \ref{subsubsection:convergence_in_a_normed_measure_space} we see that it inherits convergence in a normed measure space. In Section \ref{subsubsection:geometric_ergodicity} Theorem \ref{thm:inherited_geometric_ergodicity} we see that it inherits geometric ergodicity, and therefore we also state a CLT type result for the occlusion process in \ref{cor:CLT_from_geometric_ergodicity}.

In section \ref{section:efficient_simulation_of_the_occlusion_process} we describe a way to simulate the occlusion process defined in (\ref{eqn:K_occ_definition}) with a variational distribution $Q$ and a threaded computer, in the same amount of time it would take to run the underlying Markov chain. In Section \ref{subsubsection:sampling_from_the_P_is} we show how to simultaneously define the regions $\{\mathsf{X}_i:i\in [R]\}$ and sample from the measures $P_i$ for $i \in [R]$. In section \ref{subsubsection:implementation_and_computation_cost} we show how to run the process in an embarrassingly parallel fashion on a threaded computer, producing an algorithm at no extra time-complexity to the underlying Markov chain. We present this implementation as pseudocode in Algorithm \ref{alg:embarrassingly_parallel}.

In Section \ref{section:numerical_experiments} we have two numerical experiments which demonstrate the ability of the occlusion process to produce decorrelated samples, resulting in an estimator of reduced variance. Section \ref{subsection:bimodal_gaussian_mixture} compares the occlusion process against a Random Walk Metropolis chain on a bimodal Gaussian mixture in $d = 1$ and $d = 100$. In Section \ref{subsection:Ising_model} we compare the performances of the occlusion process, the Metropolis algorithm (Algorithm \ref{alg:metropolis}), and the Wolff algorithm (Algorithm \ref{alg:wolff}) in sampling from the Ising model (\ref{eqn:Ising_measure}) at a range of temperatures, and on various graphs. To simulate the particular form of the occlusion process in Algorithm \ref{alg:embarrassingly_parallel} we need a variational approximation to the Ising model, and so we propose one in Section \ref{subsubsection:an_efficiently_simulable_variational_approximation_to_the_Ising_model}. We show the experimental results in section \ref{subsubsection:Ising_experiment_results} and Figure \ref{fig:Ising_results}. They show that the occlusion process does decorrelate the states in the estimator, resulting in a reduced variance in the case where the states are correlated. However when the states are anticorrelated, as in the Wolff algorithm at low temperatures, the occlusion process can increase the variance.

\section{The occlusion process}\label{section:the_occlusion_process}

\subsubsection{The resolution}\label{subsubsection:the_resolution}

We first introduce a mathematical object we call the \emph{resolution}. It is an object which naturally occurs
due to the presence of the regions $\{\mathsf{X}_{i}:i\in[R]\}$. Specifically given a target $P$,
a function $f\in L^{2}(P)$, and a partition $\{\mathsf{X}_{i}:i\in[R]\}$
of $\mathsf{X}$, we define the resolution $\overrightarrow{P}f\in L^{2}(P)$
as $\overrightarrow{P}f(x):=\sum_{i=1}^{R}\mu_{i}\mathds{1}\{x\in\mathsf{X_{i}}\}$
for all $x\in\mathsf{X}$. Having introduced the resolution, it is
natural for us to define its \emph{orthogonal counterpart}: $\overleftarrow{P}f:=f-\overrightarrow{P}f$
such that $f=\overrightarrow{P}f+\overleftarrow{P}f$ for all $f\in L^{2}(P)$.
The salience of these two functions can be seen when we look at their
variances: $\textup{Var}_{P}(\overrightarrow{P}f)=\sum_{i=1}^{R}P(\mathsf{X}_{i})(\mu_{i}-\mu)^{2}$
and $\textup{Var}_{P}(\overleftarrow{P}f)=\sum_{i=1}^{R}P(\mathsf{X}_{i})\sigma_{i}^{2}$.
The variance of the resolution represents the variation between the
regions and the variance of its orthogonal counterpart represents
the variation within the regions, which is why they appear so much
in our analysis. For instance if we condition on the regions we have
that $\textup{Var}_{P}(f)=\textup{Var}_{P}(\overrightarrow{P}f)+\textup{Var}_{P}(\overleftarrow{P}f)$.
This is due to the fact that $\langle\overrightarrow{P}f,\overleftarrow{P}f\rangle_{P}=\textup{Cov}_{P}(\overrightarrow{P}f,\overleftarrow{P}f)=0$
hence the name `orthogonal counterpart'. 

\subsection{Stratified sampling}\label{subsection:stratified_sampling}

Stratified sampling \cite[Chs 5 - 5A]{cochran:1977} is the closest cousin
to the occlusion process in the field of variance reduction techniques.
It is a technique in survey sampling whereby we stratify a population
and sample from individual strata. Given that we know the populations
of the strata relative to the total population, we can combine the
samples into an estimator whose variance is reduced relative to that
of a MC estimator formed using samples of the whole population. In
the following introduction we borrow some of the notation and elementary
results from \cite[Ch 8]{owen:2013}. In our context the
strata consist of the regions $\{\mathsf{X}_{i}:i\in[R]\}$ and instead
of a population we simply have the distribution $P$. Stratified sampling
therefore assumes that we know $P(\mathsf{X}_{i})$ and can sample
from $P_{i}$ for all $i\in[R]$. Given a sample size $n_{i}\in\mathbb{N}\backslash\{0\}$
per region, we take $n_{i}$ independent samples from each $P_{i}$
and form the estimator:
\begin{equation}\label{eqn:var_hat_mu_strat}
    \hat{\mu}_{\textup{strat}}:=\sum_{i=1}^{R}\frac{P(\mathsf{X}_{i})}{n_{i}}\sum_{j=1}^{n_{i}}f(Y_{ij})
\end{equation}
where $Y_{ij}\sim P_{i}$ independently for all $j\in[N_{i}]$, for
all $i\in[R]$. This estimator is unbiased and has variance
\[
\textup{Var}(\hat{\mu}_{\textup{strat}})=\sum_{i=1}^{R}P(\mathsf{X}_{i})^{2}\frac{\sigma_{i}^{2}}{n_{i}}
\]
The idea is then to choose the $n_{i}$'s to minimise the variance,
subject to some cost constraints. There is also the larger question
of how to choose the regions.

\subsubsection{Proportional allocation}\label{subsubection:proportional_allocation}

Stratified sampling with proportional allocation has $n_{i}=P(\mathsf{X}_{i})n$
(we neglect any effects due to rounding). Using such a scheme yields
a variance of
\[
\textup{Var}(\hat{\mu}_{\textup{prop}})=\frac{1}{n}\textup{Var}_{P}(\overleftarrow{P}f)
\]
and hence $\textup{Var}(\hat{\mu}_{\textup{prop}})\leq\textup{Var}(\hat{\mu})$.
We also have a quantitive relationship between $\textup{Var}(\hat{\mu}_{\textup{prop}})$
and $\textup{Var}(\hat{\mu})$ using the resolution: $\textup{Var}(\hat{\mu}_{\textup{prop}})=(1-\textup{Corr}_{P}(f,\overrightarrow{P}f)^{2})\textup{Var}(\hat{\mu})$,
and so the more $f$ looks piecewise constant on the regions, the
smaller the ratio between $\textup{Var}(\hat{\mu}_{\textup{prop}})$
and $\textup{Var}(\hat{\mu})$. It is known that proportional allocation
is not the universally optimal scheme \citep{Neyman:1934}, since, for instance, taking
any samples from regions in which $f$ is extremely flat will do nothing
to reduce $\textup{Var}(\hat{\mu}_{\textup{strat}})$ in \ref{eqn:var_hat_mu_strat}.

\subsection{The occlusion process}\label{subsection:the_occlusion_process}

In the MCMC setting, we will not know how to sample from $P$ nor
will we know the weights $P(\mathsf{X}_{i})$ beforehand, but we can
sample from a $P$-invariant Markov chain. We can still sample from
the $P_{i}$'s, but the computational cost may be random. Specifically, we assume access to a $P$-invariant Markov chain $\{X_{t}\}_{t=1}^{n}$
with $X_{1}\sim P$. We also assume that we can sample from each $P_{i}$
during the running of the Markov chain, but the number of samples
will depend on our specific resources such as the number of threads
we have access to.


\subsubsection{The fully occluded estimator}\label{subsubsection:the_fully_occluded_estimator}

We first introduce the occluded estimator in the instance in which every sample from the Markov chain is occluded
by a sample from $P_i$. This is, in some sense, an `ideal' form
of the final estimator, since we would use it if we were able to get
sufficiently many samples from $P_i$. We make the conservative assumption
that we only use a single sample from $P_i$ sample for each sample in the
Markov chain (in the corresponding region). The estimator is then
\begin{equation}\label{eqn:mu_hat_ideal}
    \hat{\mu}_{\textup{ideal}}:=\frac{1}{n}\sum_{i=1}^{R}\sum_{j=1}^{N_{i}}f(Y_{ij})
\end{equation}

where $Y_{ij} \sim P_i$ for all $j \in [N_i]$ and $N_{i}=\sum_{t=1}^{n}\mathds{1}\{X_{t}\in\mathsf{X}_{i}\}$
is just the time the Markov chain spends in the $i$th region, for
all $i\in[R]$. We assume that $Y_{ks} \perp Y_{k's'} | \{N_i\}_{i = 1}^R$ for all possible pairs $(k, s)$ and $(k',s')$. Therefore the samples $Y_{ij}$ are independent of each other given how many of them we need to collect. We also assume that $Y_{ks} \perp \{X_t\}_{t = 1}^n | \{N_i\}_{i = 1}^R$ for all pairs $(k, s)$ and so the samples $Y_{ij}$ are independent of the Markov chain, given how long it spends in each region. 

Estimates of the weights $P(\mathsf{X}_{i})$ are naturally included via the $N_{i}$'s. The estimator is defined
equivalently to the stratified sampling with proportional allocation
estimator except that the $N_{i}$'s are selected using a Markov chain.

\begin{figure}[h]
    \centering    \includegraphics[scale = 1.25]{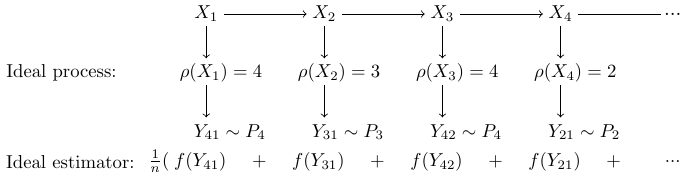}
    \caption{A DAG representing the process in which every state $X_t$ in the Markov chain is replaced by a sample from $P_{\rho(X_t)}$ in the estimator.}
    \label{fig:DAG_ideal}
\end{figure}

Figure \ref{fig:DAG_ideal} shows the process in the form of a DAG and the ensuing estimator, along with some sample regions into which the Markov chain states fall.

\begin{prop}
\label{prop:ideal_estimator}The estimator $\hat{\mu}_{\textup{ideal}}$ in (\ref{eqn:mu_hat_ideal})
is unbiased. It has variance
\begin{equation}\label{eqn:var_mu_hat_ideal}
    \textup{Var}(\hat{\mu}_{\textup{ideal}})=\frac{1}{n}\textup{Var}_{P}(\overleftarrow{P}f)+\textup{Var}\left(\frac{1}{n}\sum_{t=1}^{n}\overrightarrow{P}f(X_{t})\right)
\end{equation}
\end{prop}

Hence $\textup{Var}(\hat{\mu}_{\textup{ideal}})$ is just $\textup{Var}(\hat{\mu}_{\textup{prop}})$
plus a penalty paid for the Markovianity, albeit through the resolution
$\overrightarrow{P}f$. Proof of Proposition \ref{prop:ideal_estimator} can be found in section \ref{proof:ideal_estimator}. In the stratified sampling with proportional
allocation case we had that the variance of the estimator was smaller than the variance of the MC estimator. Interestingly, this is
not the case when we compare $\textup{Var}(\hat{\mu}_{\textup{ideal}})$
with the variance of the Markov chain estimator:
\begin{fact}\label{fact:counterexample_existence}
There exist distributions $P$, partitions $\{\mathsf{X}_{i}:i\in[R]\}$,
functions $f\in L^{2}(P)$, and $P$-invariant Markov chains $\{X_{t}\}_{t=1}^{n}$
such that $\textup{Var}(\hat{\mu}_{\textup{ideal}})>\textup{Var}(n^{-1}\sum_{t=1}^{n}f(X_{t}))$ for some $n\in \mathbb{N}\backslash\{0\}$.
\end{fact}

The fact is obtained by using the trivial partition $\{\mathsf{X}\}$
which dictates that $\overrightarrow{P}f=P(f)$ and hence that $\textup{Var}(n^{-1}\sum_{t=1}^{n}\overrightarrow{P}f(X_{t}))=0$
and $\textup{Var}_{P}(\overleftarrow{P}f)=\textup{Var}_{P}(f)$. In
this case $\textup{Var}(\hat{\mu}_{\textup{ideal}})$ is simply the
variance of the MC estimator. However, there exist scenarios in which
the variance of the Markov chain estimator is less than the variance
of the MC estimator, for instance, let $K$ be the operator associated with the Markov chain and $f\in L^2(P)$ be such that $\langle f, Kf \rangle_P < 0$. Then for $n=2$ we have that
\begin{equation}\label{eqn:witness_to_counterexample}
    \begin{split}
        \textup{Var}(n^{-1}\sum_{t = 1}^nf(X_t)) &= \frac{1}{2}\textup{Var}_P(f(X)) + \frac{1}{4}\textup{Cov}_P(f(X), Kf(X))\\
        &= \frac{1}{2}\textup{Var}_P(f(X)) + \frac{1}{4}\langle f, Kf \rangle_P - \frac{1}{4}\mathbb{E}_P[f(X)]^2 < \frac{1}{2}\textup{Var}_P(f(X))
    \end{split}
\end{equation}
See \cite[Corollary 15]{neal:2024}
or \cite[Theorem 2]{liu:2024} for asymptotic instances of the above phenomenon. It is unknown to the
authors whether Fact \ref{fact:counterexample_existence} can be extended to partitions with multiple
regions:
\begin{open_q}\label{open_question:non_trivial_partition}
Does there exist a distribution $P$, a partition $\{\mathsf{X}_{i}:i\in[R]\}$
with $R>1$, a function $f\in L^{2}(P)$, and a $P$-invariant Markov
chain $\{X_{t}\}_{t=1}^{n}$ such that $\textup{Var}(\hat{\mu}_{\textup{ideal}})>\textup{Var}(n^{-1}\sum_{t=1}^{n}f(X_{t}))$ for some $n \in \mathbb{N}\backslash \{0\}$?
\end{open_q}

\cite[Corollary 15]{neal:2024} uses antithetic Markov chains to produce
estimators with lower variances than MC estimators. An antithetic
Markov chain is one whose kernel is completely negative (apart from
an eigenvalue at 1). The witness (\ref{eqn:witness_to_counterexample}) to Fact \ref{fact:counterexample_existence} also exploits the existence of a negative part of $K$'s spectrum.

Regardless of the sign of the Markov kernel, we are able to establish
the following:
\begin{lem}
\label{lem:K_variance_dominance_over_resolution}Let $K$ be the
Markov operator associated with the chain $\{X_{t}\}_{t=1}^{n}$.
We have that 
\[
\textup{Var}\left(n^{-1}\sum_{t=1}^{n}\overrightarrow{P}f(X_{t})\right)\leq\textup{Var}\left(n^{-1}\sum_{t=1}^{n}f(X_{t})\right)
\]
for all $n \in \mathbb{N}\backslash \{0\}$.
\end{lem}

Proof of Lemma \ref{lem:K_variance_dominance_over_resolution} can be found in section \ref{proof:K_variance_dominance_over_resolution}. Clearly then if $\textup{Var}_{P}(\overleftarrow{P}f)$ is sufficiently
small, i.e. if $f\approx\overrightarrow{P}f$, we would have that
$\textup{Var}(\hat{\mu}_{\textup{ideal}})\leq\textup{Var}(n^{-1}\sum_{t=1}^{n}f(X_{t}))$.
This mirrors the case in stratified sampling with proportional allocation
where we achieve an optimal variance when $f=\overrightarrow{P}f$. Interestingly, we also have that $\textup{Var}(\hat{\mu}_{\textup{ideal}})\leq\textup{Var}(n^{-1}\sum_{t=1}^{n}f(X_{t}))$ when $f \equiv \overleftarrow{P}f + P(f)$ and the Markov chain is positive:

\begin{prop}\label{prop:ideal_variance_dominance}
    The following conditions are individually sufficient for $\textup{Var}(\hat{\mu}_{\textup{ideal}})\leq\textup{Var}(n^{-1}\sum_{t=1}^{n}f(X_{t}))$:
    \begin{enumerate}
        \item That $f(x) = \overrightarrow{P}f(x)$ for all $x \in \mathsf{X}$.
        \item That $f(x) = \overleftarrow{P}f(x) + P(f)$ for all $x \in \mathsf{X}$ and that the spectrum of $K$ is positive.
    \end{enumerate}
\end{prop}

For a proof, see Section \ref{proof:ideal_variance_dominance}. Condition 1. is equivalent to $f$ being piecewise constant over the regions. We state a natural case in which the function requirement of condition 2. holds below:

\begin{ex}\label{ex:odd_function_even_measure}
    Let $f$ be an odd function, with $f(-x) = -f(x)$ for all $x \in \mathsf{X}$, and let $P$ be even, such that $P(A) = P(-A)$ for all $A \in \mathcal{X}$. Then if the regions satisfy $\mathsf{X}_i = -\mathsf{X}_i$ for all $i \in [R]$, we have $\overrightarrow{P}f \equiv P(f) \equiv 0$ and hence that $f \equiv \overleftarrow{P}f + P(f)$.
\end{ex}

\subsubsection{The occluded estimator}\label{subsubsection:the_occluded_estimator}

Since we might not have enough samples from each $P_{i}$ we construct
the actual estimator using samples from the Markov chain, and occlude
a sample from the chain with a sample from $P_{i}$ whenever one is drawn. We extend the state space $\mathsf{X}$ of the Markov
chain to include an indicator variable $s\in\{0,1\}$ which indicates a sample from $P_{i}$ and the space of the sample
from $P_{i}$. We define $\alpha : [R] \to [0, 1]$ such that $\alpha(i)$ is the probability of a sample from $P_i$ for all $i \in [R]$. For all $(x,s,y)\in\mathsf{X}\times\{0,1\}\times\mathsf{X}$
the occlusion process is then constructed using the Markov kernel $K_{\textup{occ}}((x,s,y)\to.):\mathcal{X}\times\{0,1\}\times\mathcal{X}\to\mathbb{R}^{+}$
with
\begin{equation}\label{eqn:K_occ_definition}
    K_{\textup{occ}}((x,s,y)\to(dx',s',dy')):=K(x\to dx')\left(\alpha(\rho(x'))\mathds{1}\{s'=1\}+\left(1-\alpha(\rho(x'))\right)\mathds{1}\{s'=0\}\right)P_{\rho(x')}(dy')
\end{equation}
We use
the estimator
\begin{equation}\label{eqn:mu_hat_occ}
    \hat{\mu}_{\textup{occ}}:=\frac{1}{n}\sum_{t=1}^{n}f_{\textup{occ}}(X_{t},S_{t},Y_{t})
\end{equation}
where $f_{\textup{occ}}(X_{t},S_{t},Y_{t}):=\mathds{1}\{S_{t}=0\}f(X_{t})+\mathds{1}\{S_{t}=1\}f(Y_{t})$,
and so by construction $X_{t}$ is occluded by $Y_{t}$ whenever we
obtain a successful sample from $P_{\rho(X_{t})}$. Even though the process described above generates $Y_t$ for all $t \in [n]$, in actuality we will only need to generate $Y_t$ when $S_t = 1$ since it is only then that $Y_t$ is included in the estimator. Note that if $\alpha\equiv1$
we have that $\hat{\mu}_{\textup{occ}}$ is just $\hat{\mu}_{\textup{ideal}}$ in (\ref{eqn:mu_hat_ideal}). For a pictorial representation, see Figure (\ref{fig:DAG_diagram}).


Since $K$ is $P$-invariant we have that $K_{\textup{occ}}$ is $P_{\textup{occ}}$-invariant
where we define
\begin{equation}\label{eqn:P_occ_definition}P_{\textup{occ}}(dx,s,dy):=P(dx)\left(\alpha(\rho(x))\mathds{1}\{s=1\}+\left(1-\alpha(\rho(x))\right)\mathds{1}\{s=0\}\right)P_{\rho(x)}(dy)
\end{equation}

for all $dx\in\mathcal{X},s\in\{0,1\},dy\in\mathcal{X}$. Hence the marginal processes of the occlusion process obey the following
laws in equilibrium:
\begin{itemize}
\item The marginal law of $\{X_{t}\}$ in equilibrium is $P$.
\item The marginal law of $\{S_{t}\}$ in equilibrium is defined by the probability mass
function $\bar{\alpha}\mathds{1}\{s=1\}+(1-\bar{\alpha})\mathds{1}\{s=0\}$
where 
\[
\bar{\alpha}:=\sum_{i=1}^{R}\alpha(i)P(\mathsf{X}_{i})
\]
\item The marginal law of $\{Y_{t}\}$ in equilibrium is $P$.
\item Therefore defining $Z_t :=\mathds{1}\{S_{t}=0\}X_{t}+\mathds{1}\{S_{t}=1\}Y_{t}$ such that $f_\textup{occ}(X_t,S_t,Y_t) \equiv f(Z_t)$ we have that the marginal law of $\{Z_t\}$ in equilibrium is $P$.
\end{itemize}

Note that $P_{\textup{occ}}(f_{\textup{occ}})=\mu$ and $\textup{Var}_{P_{\textup{occ}}}(f_{\textup{occ}})=\textup{Var}_{P}(f)$.
For the bias and variance of $\hat{\mu}_{\textup{occ}}$ we have the
following:
\begin{prop}
\label{prop:variance_mu_occ}When evaluated on the process $\{(X_{t},S_{t},Y_{t})\}_{t=1}^{n}$
with $(X_{1},S_{1},Y_{1})\sim P_{\textup{occ}}$, $\hat{\mu}_{\textup{occ}}$ (\ref{eqn:mu_hat_occ})
is unbiased and has variance
\begin{equation}\label{eqn:var_mu_hat_occ}
    \textup{Var}(\hat{\mu}_{\textup{occ}})=\textup{Var}(\hat{\mu}_{\textup{ideal}})+\frac{1}{n}2\sum_{k=1}^{n-1}\frac{n-k}{n}C_{k}
\end{equation}
where
\[
C_{k}:=\textup{Cov}_{P}(f_{a}(X),K^{k}f_{a}(X))-\textup{Cov}_{P}(\overrightarrow{P}f_{a}(X),K^{k}\overrightarrow{P}f_{a}(X))
\]
with $f_{a}(x):=\left(1-\alpha(\rho(x))\right)f(x)$ for all $x\in\mathsf{X}$ and $\textup{Var}(\hat{\mu}_\textup{ideal})$ is as defined in (\ref{eqn:var_mu_hat_ideal}).
\end{prop}

Proof of the above proposition can be found in section \ref{proof:variance_mu_occ}. Hence when $\alpha \equiv 0$ we have $\textup{Var}(\hat{\mu}_{\textup{occ}})=\textup{Var}(n^{-1}\sum_{t=1}^{n}f(X_{t}))$
as expected, and when $\alpha\equiv1$ we have $\textup{Var}(\hat{\mu}_{\textup{occ}})=\textup{Var}(\hat{\mu}_{\textup{ideal}})$. In this latter $\alpha \equiv 1$ case we would have $\textup{Var}(\hat{\mu}_{\textup{occ}}) \leq \textup{Var}(n^{-1}\sum_{t=1}^{n}f(X_{t}))$  when $f$ satisfies either condition 1. or 2. from Proposition \ref{prop:ideal_variance_dominance}. These conditions may seem improbable, but in section \ref{subsection:Ising_model} we see a practical instantiation of condition 2. in the form outlined in Example \ref{ex:odd_function_even_measure} where the function $f$ is odd and the target measure $P$ is even. See Figure \ref{fig:Ising_results} for the concomitant reductions of variance, and Section \ref{subsubsection:satisfaction_of_theoretical_conditions_for_variance_reduction} for a justification of why in this case, we satisfy condition 2.

\section{Inherited theoretical properties of the occlusion process}\label{section:inherited_theoretical_properties_of_the_occlusion_process}

\subsubsection{Basic properties}\label{subsubsection:basic_properties}

Since the occlusion process $\{(X_{t},S_{t},Y_{t})\}_{t=1}^{n}$ is
derived from an underlying Markov chain $\{X_{t}\}_{t=1}^{n}$, we
would like the process to inherit the `good' properties of the Markov
chain when they exist. For example we have the following inheritances:
\begin{prop}
\label{prop:inherited_reversibility}If $K$ is $P$-reversible then
$K_{\textup{occ}}$ is $P_{\textup{occ}}$-reversible.
\end{prop}
A proof of Proposition \ref{prop:inherited_reversibility} can be found in section \ref{proof:inherited_reversibility}
\begin{prop}
\label{prop:inherited_l2}If $f$ is in $L^{2}(P)$ then $f_{\textup{occ}}$
is in $L^{2}(P_{\textup{occ}})$.
\end{prop}

Proof can be found in section \ref{proof:inherited_l2}.

\subsection{Law of Large Numbers}\label{subsection:law_of_large_numbers}

The first non-basic result that the occlusion process inherits from the Markov chain is a Law of Large Numbers (LLN). Since no quantitative rates are involved in LLNs we can not directly compare LLNs on the Markov chain with LLNs on the occlusion process but the inheritance result stands nonetheless.

\begin{thm}\label{thm:LLN_inheritance}
  The following are equivalent:
  \begin{enumerate}
      \item For all probability measures $\mu$ on $\mathcal{X}$ and measurable functions $g:\mathsf{X} \to \mathbb{R}$ such that $g \in L^1(P)$ we have that
      \[
      \lim_{n \to \infty}n^{-1}\sum_{t = 1}^n g(X_t) = P(g)
      \]
      almost surely with $X_1 \sim \mu$.
      \item For all probability measures $\mu_\textup{occ}$ on $\mathcal{X} \times 2 ^ {\{0, 1\}} \times \mathcal{X}$ and for all measurable functions $g_\textup{occ}: \mathsf{X} \times \{0, 1\} \times \mathsf{X}$ such that $g_\textup{occ} \in L^1(P_\textup{occ})$ we have that
      \[
      \lim_{n \to \infty} n^{-1}\sum_{t=1}^n g_\textup{occ}(X_t, S_t, Y_t) = P_\textup{occ}(g_\textup{occ})
      \]
      almost surely with $(X_1, S_1, Y_1) \sim \mu_\textup{occ}$.
  \end{enumerate}
\end{thm}

The proof in section \ref{proof:LLN_inheritance} uses \cite[Proposition 3.5]{douc:2023}.

\subsection{Convergence in a given norm}\label{subsection:convergence_in_a_given_norm}

\subsubsection{Convergence in a normed function space}\label{subsubsection:convergence_in_a_normed_function_space}

One way to measure the efficiency of an MCMC algorithm with kernel
$K$ is to compare $K^{t}$ to its equilibrium distribution via their
action on measurable functions. The outputs of these actions are themselves
functions, so to compare we need to use a norm defined on the appropriate
function space.
\begin{defn}\label{defn:function_normed_convergence}
A Markov chain with kernel $K$ converges to a distribution $P$ in
a normed function space $(\mathsf{F},\|.\|)$ with rate function $r:\mathbb{N}\backslash\{0\}\to\mathbb{R}^{+}$
when
\[
\|K^{t}f-P(f)\|\leq C_{f}r(t)
\]
for all $f\in\mathsf{F}$ and $t\in\mathbb{N}\backslash\{0\}$, where
$C_{f}>0$ is a constant that depends on $f$ and $r(t)\searrow0$.
\end{defn}

Often we have that $C_{f}=C\|f-P(f)\|$
with $C>0$. For the occlusion process $\{(X_{t},S_{t},Y_{t})\}_{t=1}^{n}$
to inherit convergence in a normed space of $\{X_{t}\}_{t=1}^{n}$
we need some way to relate the normed spaces that $K$ acts on to
the normed spaces that $K_{\textup{occ}}$ acts on. Given a normed
function space $(\mathsf{F},\|.\|)$ of functions on $\mathsf{X}$,
let $\mathsf{F}_{\textup{occ}}$ be the vector space of measurable
functions of the form $g(x,s,y)=\mathds{1}\{s=0\}f(x)+\mathds{1}\{s=1\}f(y)$
on $\mathsf{X}\times\{0,1\}\times\mathsf{X}$ into $\mathbb{R}$ where
$f\in(\mathsf{F},\|.\|)$. These are the functions that $K_{\textup{occ}}$
will act on to produce the occlusion process. We define the class
of normed spaces
\[
\mathcal{C}_{\|.\|}:=\{(\mathsf{G},\|.\|_{\mathsf{G}}):\mathsf{G\subseteq\mathsf{F}_{\textup{occ}}},\|g\|_{\mathsf{G}}=\|g\|\textup{ when }g\textup{ is a function of its first argument only}\}
\]
For instance, if $\|.\|$ is the sup norm on functions from $\mathsf{X}\to\mathbb{R}$
then $\mathcal{C}_{\|.\|}$will contain the normed spaces with the
sup norm on functions from $\mathsf{F}_{\textup{occ}}$. Equipped
with this definition, we have the following inheritance
\begin{thm}
\label{thm:inherited_normed_function_convergence}Say the Markov chain
$\{X_{t}\}_{t=1}^{n}$ converges to $P$ in the normed function space
$(\mathsf{F},\|.\|)$ with rate function $r(t)$ and constant $C_{f}$.
Then for all normed function spaces $(\mathsf{G},\|.\|_{\mathsf{G}})\in\mathcal{C}_{\|.\|}$
and for all functions $g(x,s,y)=\mathds{1}\{s=0\}f(x)+\mathds{1}\{s=1\}f(y)\in(\mathsf{G},\|.\|_{\mathsf{G}})$
we have that
\[
\|K_{\textup{occ}}^{t}g-P_{\textup{occ}}(g)\|_{\mathsf{G}}\leq C_{f_{\alpha}}r(t)
\]
where
\[
f_{\alpha}(x):=\left(1-\alpha(\rho(x))\right)f(x)+\alpha(\rho(x))\overrightarrow{P}f(x)
\]
$K_{\textup{occ}}$ is as defined in (\ref{eqn:K_occ_definition}), and $P_{\textup{occ}}$ is as defined in (\ref{eqn:P_occ_definition})\end{thm}

The proof relies on the fact that for a given function $g\in(\mathsf{G},\|.\|_{\mathsf{G}})$
with $(\mathsf{G},\|.\|_{\mathsf{G}})\in\mathcal{C}_{\|.\|}$ we have
$K_{\textup{occ}}^{t}g=K^{t}f_{\alpha}$, see section \ref{proof:inherited_normed_function_convergence}. Because of this fact, we
also inherit lower bounds on the convergence of the occlusion process: 
\begin{prop}
\label{prop:inherited_normed_function_bounds}Say the convergence
of the Markov chain $\{X_{t}\}_{t=1}^{n}$ to $P$ in the normed function
space $(\mathsf{F},\|.\|)$ is lower bounded as follows:
\[
\|K^{t}f-P(f)\|\geq C_{f}^{\dagger}r^{\dagger}(t)
\]
for some $C_{f}^{\dagger}>0$ and $r^{\dagger}:\mathbb{N}\backslash\{0\}\to\mathbb{R}^{+}$
and for all $f\in(\mathsf{F},\|.\|)$. Then for all normed function
spaces $(\mathsf{G},\|.\|_{\mathsf{G}})\in\mathcal{C}_{\|.\|}$ and
for all functions $g(x,s,y)=\mathds{1}\{s=0\}f(x)+\mathds{1}\{s=1\}f(y)\in(\mathsf{G},\|.\|_{\mathsf{G}})$
we have that
\[
\|K_{\textup{occ}}^{t}g-P_{\textup{occ}}(g)\|_{\mathsf{G}}\geq C_{f_{\alpha}}^{\dagger}r^{\dagger}(t)
\]
where
\[
f_{\alpha}(x):=\left(1-\alpha(\rho(x))\right)f(x)+\alpha(\rho(x))\overrightarrow{P}f(x)
\]
$K_{\textup{occ}}$ is as defined in (\ref{eqn:K_occ_definition}), and $P_{\textup{occ}}$ is as defined in (\ref{eqn:P_occ_definition})
\end{prop}

If $C_{f_{\alpha}}\leq C_{f}$ or $C_{f_{\alpha}}^\dagger\leq C_{f}^\dagger$ we could have better convergence of
the occlusion process as compared with the base Markov chain it sits
upon. The following example shows such a case.
\begin{ex}
\label{ex:sup_norm}Say $(\mathsf{F},\|.\|)$ is the space of bounded
continuous functions with the sup norm, and that $\{X_{t}\}_{t=1}^{n}$
converges to $P$ in $(\mathsf{F},\|.\|)$ with rate function $r(t)$
and constant $C_{f}:=C\|f-P(f)\|$ with $C>0$. Then the occlusion
process $\{(X_{t},S_{t},Y_{t})\}_{t=1}^{n}$ converges to $P_{\textup{occ}}$
in all spaces $(\mathsf{G},\|.\|_{\mathsf{G}})$ in $\mathcal{C}_{\|.\|}$
with rate function $r(t)$ and constant $C_{f_{\alpha}}:=C\|f_{\alpha}-P(f_{\alpha})\|$.
Note that $C_{f_{\alpha}}\leq C_{f}$. For a proof of this fact see
\ref{proof:sup_norm}. The same is true for any similarly defined
constants $C_{f_{\alpha}}^{\dagger}$ and $C_{f}^{\dagger}$ in lower
bounds on convergence, for exactly the same reasons.
\end{ex}

\subsubsection{Convergence in a normed measure space}\label{subsubsection:convergence_in_a_normed_measure_space}

Another way to examine the efficiency of an MCMC algorithm with kernel
$K$ is to establish bounds on the distance between $\mu K^{t}$ and $P$
in some normed measure space.
\begin{defn}\label{defn:measure_normed_convergence}
A Markov chain with kernel $K$ converges to a distribution $P$ in
a normed measure space $(\mathsf{M},\|.\|)$ with rate function $r(t):\mathbb{N}\backslash\{0\}\to\mathbb{R}^{+}$
when
\[
\|\mu K^{t}-P\|\leq C_{\mu}r(t)
\]
for all measures $\mu\in(\mathsf{M},\|.\|)$ and $t\in\mathbb{N}\backslash\{0\}$,
where $C_{\mu}>0$ is a constant that depends on $\mu$
\end{defn}

For the above definition to work, we need that measures in $(\mathsf{M},\|.\|)$
have the same state space as $K$. Again we would ideally have that
$r(t)\searrow0$. The role of the norm in the above definition is
to provide the distance between $\mu K^{t}$ and $P$. So given a
distance, we could equivalently state the convergence result without
a norm. One class of distances we could use is the class of integral
probability metrics (IPMs). It is defined using a function space $\mathsf{F}$
as follows:
\begin{defn}\label{defn:IPM}
The integral probability metric $D_{\mathsf{F}}$ defined by a function
space $\mathsf{F}$ between measures $P$ and $Q$ is defined as
\[
D_{\mathsf{F}}(P,Q):=\sup_{f\in\mathsf{F}}\left|P(f)-Q(f)\right|
\]
\end{defn}

Examples of IPMs include the Wasserstein-1 distance and the total
variation distance. As in the previous section, we define $\mathcal{C}:=\{\mathsf{G}:\textup{ for all }g\in\mathsf{G}\textup{ we have }g(x,s,y)=\mathds{1}\{s=0\}f(x)+\mathds{1}\{s=1\}f(y)\textup{ with }f\in\mathsf{F}\}$
as the class of function spaces whose members the occlusion process
admits in its estimator. Equipped with these definitions we have the
following inheritance result:
\begin{thm}
\label{thm:inherited_IPM_convergence}Say the Markov chain $\{X_{t}\}_{t=1}^{n}$
starting at $X_{1}\sim\mu$ converges to $P$ in the integral probability
metric defined over $\mathsf{F}$ with rate function $r(t)$ and constant
$C_{\mu}$. Then the occlusion process $\{(X_{t},S_{t},Y_{t})\}_{t=1}^{n}$
with $X_{1}\sim\mu$ converges to $P_{\textup{occ}}$ (\ref{eqn:P_occ_definition}) in the integral
probability metrics defined over the members of $\mathcal{C}$ with
rate function $r(t)$ and constant $C_{\mu}$ as soon as
\[
f_{\alpha}(x):=\left(1-\alpha(\rho(x))\right)f(x)+\alpha(\rho(x))\overrightarrow{P}f(x)
\]
is in $\mathsf{F}$.
\end{thm}
A proof of the above is found in section \ref{proof:inherited_IPM_convergence}.
\begin{ex}\label{ex:total_variation}
Take $\mathsf{F}=\{f:\mathsf{X}\to[0,1]\}$. Then $D_{\mathsf{F}}$
is the total variation distance. That $f_{\alpha}$ is in $\mathsf{F}$
is clear since it is the convex combination of two functions in $\mathsf{F}$.
Therefore if $\{X_{t}\}_{t=1}^{n}$ converges to $P$ in $D_{\mathsf{F}}$
the above result holds and we get convergence in total variation of
the occlusion process with the same rate.
\end{ex}
That the occlusion process inherits convergence in IPMs is an example
of a wider class of inheritance results: we include it here as an example.
For a more general result on the inheritance of convergence in normed
measure spaces see Appendix \ref{appendix:normed_measure_convergence} Theorem \ref{thm:inherited_normed_measure_convergence}.

\subsubsection{Geometric ergodicity}\label{subsubsection:geometric_ergodicity}

Geometric ergodicity of a Markov chain is the property that the distribution
of the final state in the chain tends towards the target distribution
at a geometric rate, up to a constant which depends on the starting
position of the chain. It is a particular type of convergence in a
normed measure space, where the norm is the total variation norm,
and $\mu$ is a point mass.
\begin{defn}
\label{defn:geometric_ergodicity}A $P$-invariant Markov chain generated
by a kernel $K(x\to.):\mathcal{X}\to\mathbb{R}^{+}$ is geometrically
ergodic when
\[
\|K^{t}(x\to.)-P(.)\|_{TV}\leq c(x)\exp(-\lambda t)
\]
for all $t\in\mathbb{N}\backslash\{0\}$ and $x\in\mathsf{X}$, and
for some $c(x)\in(0,\infty)$, $\lambda>0$. Here $K^{t}(x\to.):\mathcal{X}\to\mathbb{R}^{+}$
is the $t$-step kernel, and $\|.\|_{TV}$ is the total variation
norm.
\end{defn}
When a functional $f:\mathsf{X}\to\mathbb{R}$ is averaged over a
geometrically ergodic chain, it only takes a small amount of additional
work to establish a central limit theorem (CLT). For instance \cite[Theorem 2]{chan:1994} show that when $f\in L^{2+\epsilon}(P)$
is averaged over a geometrically ergodic chain, for some $\epsilon>0$,
we are guaranteed a CLT. \cite[Corollary 2.1]{roberts:97}
show that the same is true for $f\in L^{2}(P)$ when the chain is
$P$-reversible.
\begin{thm}
\label{thm:inherited_geometric_ergodicity}When the chain $\{X_{t}\}_{t=1}^{n}$
generated by $K$ is geometrically ergodic, the occlusion process
$\{(X_{t},S_{t},Y_{t})\}_{t=1}^{n}$ generated by $K_{\textup{occ}}$ (\ref{eqn:K_occ_definition})
is geometrically ergodic.
\end{thm}
This result is also a corollary of Theorem \ref{thm:inherited_IPM_convergence}, however it also stands on its own, see section \ref{proof:inherited_geometric_ergodicity} for the proof.
\begin{cor}
\label{cor:CLT_from_geometric_ergodicity}When the chain generated
by $K$ is geometrically ergodic and $P$-reversible and when $f\in L^{2}(P)$,
$\hat{\mu}_{\textup{occ}}$ admits the following CLT:
\[
\sqrt{n}\left(\hat{\mu}_{\textup{occ}}-\mu\right)\overset{d}{\to}N(0,\sigma_{\textup{occ}}^{2})\text{ as }n\to\infty
\]
where 
\[
\sigma_{\textup{occ}}^{2}:=\lim_{n\to\infty}n\textup{Var}(\hat{\mu}_{\textup{occ}})=\textup{Var}_{P_{\textup{occ}}}(f_{\textup{occ}})+2\sum_{k=1}^{\infty}\textup{Cov}_{P_{\textup{occ}}}(f_{\textup{occ}},K^{k}f_{\textup{occ}})<\infty
\]
\end{cor}
Proof of the above can be found in section \ref{proof:CLT_from_geometric_ergodicity}. It would be desirable for the occlusion process to inherit a CLT type result from a CLT in the Markov chain $\{X_t\}_{t = 1}^n$. However to establish such a result would be to establish necessary conditions for a Markov chain CLT which don't currently exist in the literature.
\begin{prop}\label{prop:asymptotic_occlusion_variance_equivalence}
When the chain generated by $K$ is geometrically ergodic and $P$-reversible, and when $f \in L^2(P)$ we have
that 
\begin{align*}
\lim_{n\to\infty}n\textup{Var}(\hat{\mu}_{\textup{occ}}) & =\textup{Var}_{P_{\textup{occ}}}(f_{\textup{occ}})+2\sum_{k=1}^{\infty}\textup{Cov}_{P_{\textup{occ}}}(f_{\textup{occ}},K^{k}f_{\textup{occ}})\\
 & =\textup{Var}_{P}(f)+2\sum_{k=1}^{\infty}\left(\textup{Cov}_{P}(\overrightarrow{P}f(X),K^{k}\overrightarrow{P}f(X))+C_{k}\right)<\infty
\end{align*}
where 
\[
C_{k}:=\textup{Cov}_{P}(f_{a}(X),K^{k}f_{a}(X))-\textup{Cov}_{P}(\overrightarrow{P}f_{a}(X),K^{k}\overrightarrow{P}f_{a}(X))
\]
with $f_{a}(x):=\left(1-\alpha(\rho(x))\right)f(x)$ for all $x\in\mathsf{X}$.
\end{prop}

Proof of the above proposition may be found in Section \ref{proof:asymptotic_occlusion_variance_equivalence}.

\section{Efficient simulation of the occlusion process}\label{section:efficient_simulation_of_the_occlusion_process}

We now describe exactly how one can simulate the occlusion process generated by the kernel in (\ref{eqn:K_occ_definition}). We show that given access to a multi-threaded computer and a variational approximation $Q$ we can calculate $\hat{\mu}_\textup{occ}$ (\ref{eqn:mu_hat_occ}) at no additional computational cost to calculating $n^{-1}\sum_{t=1}^nf(X_t)$. However due to this requirement, we cede control over the values of the $\alpha(i)$'s.

\subsection{Sampling from the $P_i$'s and defining the regions}\label{subsubsection:sampling_from_the_P_is}

If we are using MCMC to approximate samples from $P$, sampling from $P$ restricted to the regions $\mathsf{X}_1,\mathsf{X}_2,...$ will be a non-trivial task.

The trick we employ is to choose our sampling mechanism, then define the regions such that the samples from $P$ restricted to them is somehow guaranteed. Say we have access to $Q$: an easy to sample from distributional approximation to $P$. Let $Y$ be the result of the following sampling mechanism: given an arbitrary constant $C > 0$ and oracle access to the unnormalised Radon-Nikodym derivative $d\tilde{P}/d\tilde{Q}$ between $P$ and $Q$
\begin{enumerate}
    \item Sample $Y \sim Q$ and $U \sim \textup{Unif}[0, 1]$ independently of each other.
    \item If
    \[
    U \leq \frac{1}{C}\frac{d\tilde{P}}{d\tilde{Q}}(Y)\textup{ and }\frac{1}{C}\frac{d\tilde{P}}{d\tilde{Q}}(Y) \leq 1
    \]
    output $Y$, otherwise go to step 1.
\end{enumerate}
Then if we call $\mathsf{X}_C$ the region $\{y \in \mathsf{X}:\frac{1}{C}d\tilde{P}/d\tilde{Q}(y) \leq 1\}$ we have that $Y$ sampled according to the mechanism above is distributed according to $P$ restricted to $\mathsf{X}_C$, see section \ref{proof:Y_from_P_restricted} for proof. This mechanism has its roots in \cite[Section 2.3.4]{tierney:1994} and it is also used in \citep{douc:2023}.

To form the partition of $\mathsf{X}$ we do as follows: take $0 =: C_0 < C_1 < C_2 < \cdots < C_{R - 1} < C_R := \infty$ and define $\mathsf{X}_i := \{x:d\tilde{P}/d\tilde{Q}(x) \in [C_{i - 1}, C_i)\}$. Then to collect samples from the $P_i$'s we simply execute step 1 of the above mechanism, check which $\mathsf{X}_i$ $Y$ is in using the Radon-Nikodym derivative, and then use the appropriate $C_i$ in step 2. The particular $P_i$ sampled from will therefore be random and if $Y$ falls in $\mathsf{X}_R$ it is automatically rejected. The probability of a sample from $P_i$ in a single iteration is then
\begin{equation}
    \mathbb{P}_{Y \sim Q}\left(U\leq \frac{1}{C_i}\frac{d\tilde{P}}{d\tilde{Q}}(Y)\cap Y \in \mathsf{X}_i\right) = \frac{Z_P}{Z_Q}\frac{1}{C_i}P(\mathsf{X_i})
\end{equation}
where $Z_P$ and $Z_Q$ are the normalising constants of $\tilde{P}$ and $\tilde{Q}$ respectively.

\subsection{Implementation and computational cost}\label{subsubsection:implementation_and_computation_cost}

\subsubsection{Strategy for a general target $P$}\label{paragraph:general_strategy} In general we dedicate a single thread of compute to sampling the Markov chain $\{X_t\}_{t=1}^n$, and the rest of the available threads to sampling from the $P_i$'s. A straightforward way to do this is by having these final threads work in an embarrassingly parallel fashion, repeatedly iterating through the steps detailed in section \ref{subsubsection:sampling_from_the_P_is}. We stop all the threads upon some condition e.g. $\{X_t\}$ has reached a given length. The fact that the threads work in an embarrassingly parallel fashion minimises communication costs, and therefore maximises the amount of compute going into sampling from the $P_i$'s. It also minimises the amount of programmer time since in general it is far easier to implement embarrassingly parallel code than code in which the threads communicate.

At the end of the procedure we then have the Markov chain $\{X_t\}_{t=1}^n$, the list of regions it visits $\{\rho(X_t)\}_{t=1}^n$, and $\{Y_{ij}\}_{j=1}^{N_i}$ for all $i \in [R]$ where $N_i$ is the number of samples from $P_i$. We may possibly have $N_i = 0$ for some $i$, and we may even have $N_1 + \cdots + N_R > n$. We then use some procedure to assign $Y_{ij}$'s to the $X_t$'s we wish to occlude, which defines the $\alpha(\rho(x))$ term in (\ref{eqn:K_occ_definition}).

Define $T_i$ as the amount of time the Markov chain spends in region $i$ and $I_i \subseteq \{X_t\}_{t=1}^n$ the set of states in that region. In our numerical experiments for each $i \in [R]$ we assign $\min\{N_i, T_i\}$ samples from $\{Y_{ij}\}_{j=1}^{N_i}$ to occlude a uniformly sampled size $\min\{N_i, T_i\}$ subset of $I_i$. Therefore $\alpha(i) := \mathbb{E}[\min\{1,N_i / T_i\}]$ for all $i \in [R]$ where the expectation is over the randomness of the entire process. This is equivalent in distribution to occluding at each step of the process with probability $\min\{1,N_i / T_i\}$ although it is more efficient since it uses as many samples from $\{Y_{ij}\}_{j=1}^{N_i}$ as possible, for each $i \in [R]$.

See Algorithm \ref{alg:embarrassingly_parallel} for pseudocode of the whole procedure.

\begin{algorithm}
\caption{Embarrassingly parallel occlusion process}\label{alg:embarrassingly_parallel}
\SetKwInOut{Input}{input}\SetKwInOut{Output}{output}
\Input{Chain length $n$, Initial state $X_0$, $1 + C_\textup{rej}$ computational threads, Variational distribution $Q$, Constants $0 =: C_0 < C_1 < C_2 < \cdots < C_{R - 1} < C_R := \infty$}
\Output{A $P$-invariant Markov chain $\{X_t\}_{t=1}^n$, the regions it visits $\{\rho(X_t)\}_{t=1}^n$, collections of samples $\{\{Y_{ij}\}_{j = 1}^{N_i}\}_{i=1}^R$ with $Y_{ij} \sim P_i$ for all $j \in [N_i]$ and $i \in [R]$}
\textbf{Markov chain}\\
In thread 1:\\
\For{$t \in \{1,...,n\}$}{
  \begin{enumerate}
      \item Sample $X_t \sim K(X_{t - 1} \to .)$
      \item Store $X_t$ along with its region $\rho(X_t)$
  \end{enumerate}
}
\textbf{Rejection samplers}\\
In threads $j \in [C_\textup{rej}]$ concurrently:\\
\For{$j \in [C_\textup{rej}]$}{
    \begin{enumerate}
        \item Sample $Y \sim Q$ and $U \sim \textup{Unif}[0, 1]$ independently
        \item Determine the constant to use in the rejection sampler:
        \[
        C_{\rho(Y)} := \inf_{i \in [R]}\left\{C_i: \frac{d\tilde{P}}{d\tilde{Q}}(Y) < C_i\right\}
        \]
        and determine the region $i:=\rho(Y)$ that $Y$ is in.
        \item If
        \[
        U \leq \frac{1}{C_i}\frac{d\tilde{P}}{d\tilde{Q}}(Y)
        \]
        then append the sample to $\{Y_{ij}\}_{j = 1}^{N_{i}}$ (such that $N_i \leftarrow N_i + 1$), otherwise go to step 1.
    \end{enumerate}
}
\textbf{Postprocessing}\\
\Input{A $P$-invariant Markov chain $\{X_t\}_{t=1}^n$, the regions it visits $\{\rho(X_t)\}_{t=1}^n$, collections of samples $\{\{Y_{ij}\}_{j = 1}^{N_i}\}_{i=1}^R$ with $Y_{ij} \sim P_i$ for all $j \in [N_i]$ and $i \in [R]$}
\Output{The occlusion estimator $\hat{\mu}_\textup{occ}$ defined in (\ref{eqn:mu_hat_occ})}
Initialise the indicator sequence $\{S_t\}_{t = 1}^n \leftarrow \{0\}_{t = 1}^n$ and the sequence $\{Y_t\}_{t=1}^n \leftarrow \{\textup{NaN}\}_{t=1}^n$\\
\For{$i \in [R]$}{
    \begin{enumerate}
        \item Get the set of times the Markov chain $\{X_t\}_{t=1}^n$ was in region $i$:
        \[
        \mathcal{T}_i := \{t: X_t \in \mathsf{X}_i, t \in [n]\}
        \]
        and the amount of time $T_i := |\mathcal{T}_i|$.
        \item if $N_i \geq T_i$ set $S_t \leftarrow 1$ and $Y_t \leftarrow Y_{it}$ for all $t \in \mathcal{T}_i$.
        \item else:
        \begin{enumerate}
            \item Sample a subset $\mathcal{T}'_i \subseteq \mathcal{T}_i$ uniformly from the subsets of size $N_i$ in $\mathcal{T}_i$.
            \item Set $S_t \leftarrow 1$ and $Y_t \leftarrow Y_{it}$ for all $t \in \mathcal{T}'_i$.
        \end{enumerate}
    \end{enumerate}
}
Output
\[
\hat{\mu}_{\textup{occ}}=\frac{1}{n}\sum_{t=1}^{n}f_{\textup{occ}}(X_{t},S_{t},Y_{t})
\]
where $f_{\textup{occ}}(X_{t},S_{t},Y_{t}):=\mathds{1}\{S_{t}=0\}f(X_{t})+\mathds{1}\{S_{t}=1\}f(Y_{t})$
\end{algorithm}

\subsubsection{Approaches tailored to particular targets $P$}\label{paragraph:tailored_approaches} The nature of the sampling problem may offer more efficient alternatives to the general purpose strategy detailed above. Sampling from $K(x\to .)$ may be slow. For instance $P$ may be a Bayesian posterior distribution whose likelihood evaluations necessitate the solution of a system of differential equations, see \cite[section 5.1]{ma:2019}. Or perhaps $K(x \to .)$ is the kernel of a Metropolis-Hastings algorithm whose proposal distribution takes a while to sample from e.g. preconditioned Hamiltonian Monte Carlo in high dimension. In this regime, in the time it takes $K$ to move from a given $X_t$ to $X_{t + 1}$ we could concentrate the parallel computational resources to sampling from $P_{\rho(X_t)}$. This however incurs some communication costs between threads.

In another scenario, it may be that the regions are defined naturally given the state space. For example it may be that $\mathsf{X}$ can be written as the disjoint union of $R$ connected sets. Then we could have a variational distribution $Q_i$ for each region $\mathsf{X}_i$.

\subsection{Choice of regions}\label{subsubsection:choice_of_regions}

The way that the regions $\{\mathsf{X}_{i}:i\in[R]\}$ influence the statistical properties of the occlusion process is via the probabilities of taking sucessful rejection samples $\alpha(i)$ for $i\in[R]$ and the amount of time the Markov chain $\{X_t\}_{t=1}^n$ spends in each region. As stated at the end of section \ref{subsubsection:the_occluded_estimator}, if the regions are such that $\alpha \equiv 1$ i.e. each $P_i$ is easy to sample from, we would have $\textup{Var}(\hat{\mu}_{\textup{occ}})=\textup{Var}(\hat{\mu}_\textup{ideal})$. In the opposite scenario, if the regions are such that $\alpha \equiv 0$ i.e. each $P_i$ is impossibly hard to sample from, we would have $\textup{Var}(\hat{\mu}_{\textup{occ}})=\textup{Var}(n^{-1}\sum_{t=1}^{n}f(X_{t}))$.

It is difficult to tell from the form of (\ref{eqn:var_mu_hat_occ}) how exactly $\textup{Var}(\hat{\mu}_{\textup{occ}})$ will vary with each individual $\alpha(i)$, but as stated below Lemma \ref{lem:K_variance_dominance_over_resolution} we should choose the regions to have $f = \overrightarrow{P}f$ since such a condition ensures that $\textup{Var}(\hat{\mu}_{\textup{occ}}) = \textup{Var}(\hat{\mu}_\textup{ideal}) \leq \textup{Var}(n^{-1}\sum_{t=1}^{n}f(X_{t}))$.

With the general strategy explained in section \ref{subsubsection:implementation_and_computation_cost} we assume that we use $C_\textup{rej}\in\mathbb{N}\backslash \{0\}$ threads for the rejection sampling component, and that each thread can propose a single rejection sample per step in the Markov chain $\{X_t\}_{n = 1}^t$. Therefore we have $nC_\textup{rej}$ rejection sample attempts in total and hence $\alpha(i)=\mathbb{E}[\min\{1, N_i / T_i\}]$ where
\[
N_i\sim\textup{Binomial}\left(nC_\textup{rej}, \frac{Z_P}{Z_Q}\frac{1}{C_i}P(\mathsf{X_i})\right)
\]
and each $T_i$ will have expectation $nP(\mathsf{X}_i)$ for all $i \in [R]$. Therefore each $\alpha(i)$ will depend on the length of the Markov chain $n$.

\section{Numerical experiments}\label{section:numerical_experiments}

\subsection{Bimodal Gaussian Mixture}\label{subsection:bimodal_gaussian_mixture}

As a first demonstration of the behaviour of the occlusion process we look at sampling from a bimodal Gaussian mixture
\begin{equation}\label{eqn:bimodal_gaussian_density}
    P(dx) := (1-p)\times\mathcal{N}(x;0, \mathbf{I}_d)dx + p\times \mathcal{N}(x; m, \sigma^2 \mathbf{I}_d)dx
\end{equation}
for $p \in [0, 1]$, $m \in \mathbb{R}^d$, and $\sigma^2 > 0$. Say we are given a variational distribution $Q$ that approximates well the first component of the mixture. If we used such a distribution to do inference, then our estimates would be biased by the fact that the distribution ignores the second component. For a $\sigma^2$ which is sufficiently small the Radon-Nikodym derivative $dP/dQ(x)$ will have a large upper bound. This would be prohibitive against doing vanilla rejection sampling with $Q$ as the proposal distribution. Therefore to resolve these issues, we use the occlusion process so as to take advantage of the variational distribution and the unbiasedness of an underlying Markov chain targeting $P$.

\subsubsection{Experiment setup}

\begin{figure}
    \centering
    \includegraphics[scale = 1]{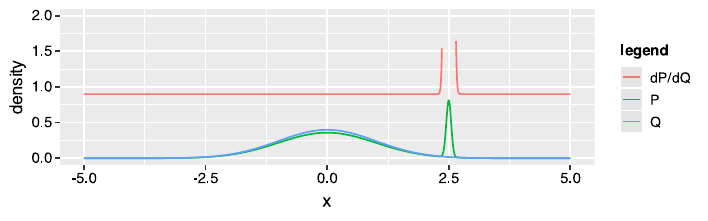}
    \caption{the target density ($P$ in the legend), the variational density ($Q$ in the legend), and the Radon-Nikodym derivative ($dP / dQ$ in the legend) for the bimodal Gaussian example in $d = 1$.}
    \label{fig:target_and_variational_density_bimodal_gaussian}
\end{figure}

We perform two experiments, one with $d = 1$ and the other with $d = 100$. The mean $m$ of the second component has all its elements set to zero, apart from its first which we set to $2.5$. We set $\sigma^2 = 0.05$ and $p = 0.1$. We set the variational distribution to be the Laplace approximation where we find the mode using gradient descent on the negative log-density of the target. In every case this resulted in $Q$ being a standard normal. For a plot of the target density, variational density, and the Radon-Nikodym derivative in $d = 1$ see Figure \ref{fig:target_and_variational_density_bimodal_gaussian}. Note that the Radon-Nikodym derivative is mostly flat apart from around $x = 2.5$ where it becomes large.

We use $R = 2$ regions, defining the first region as $\{x : dP/dQ(x) \leq 1\}$ and the second region as the complement of the first. From Figure \ref{fig:target_and_variational_density_bimodal_gaussian} we can see that the first region encompasses most of the state space, apart from a small area around the mean of the second component of the target. The number of total threads we use is 7, and we employ them in the fashion according to Section \ref{paragraph:general_strategy}, thereby having $C_\textup{rej} = 6$ threads for the rejection samplers, and one thread for the underlying Markov chain. For this underlying chain we use Random Walk Metropolis (RWM), for which we set the step size to $2.38 / \sqrt{d}$ as recommended in \citep{roberts:2001}. In each case we run the sampler for 20 seconds.

\subsubsection{Experiment results}

\begin{figure}[h]
    \centering
    \includegraphics[scale = 0.11]{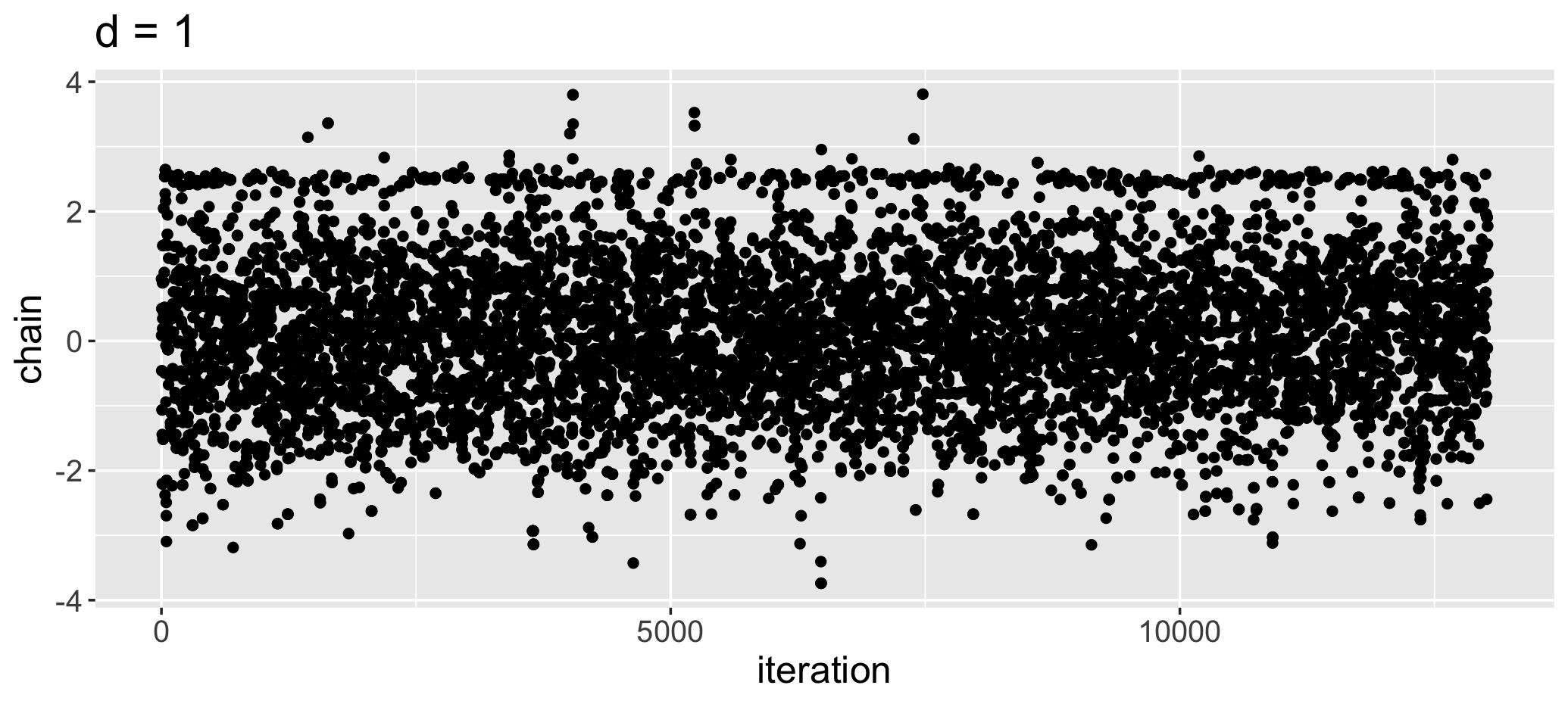}
    \includegraphics[scale = 0.11]{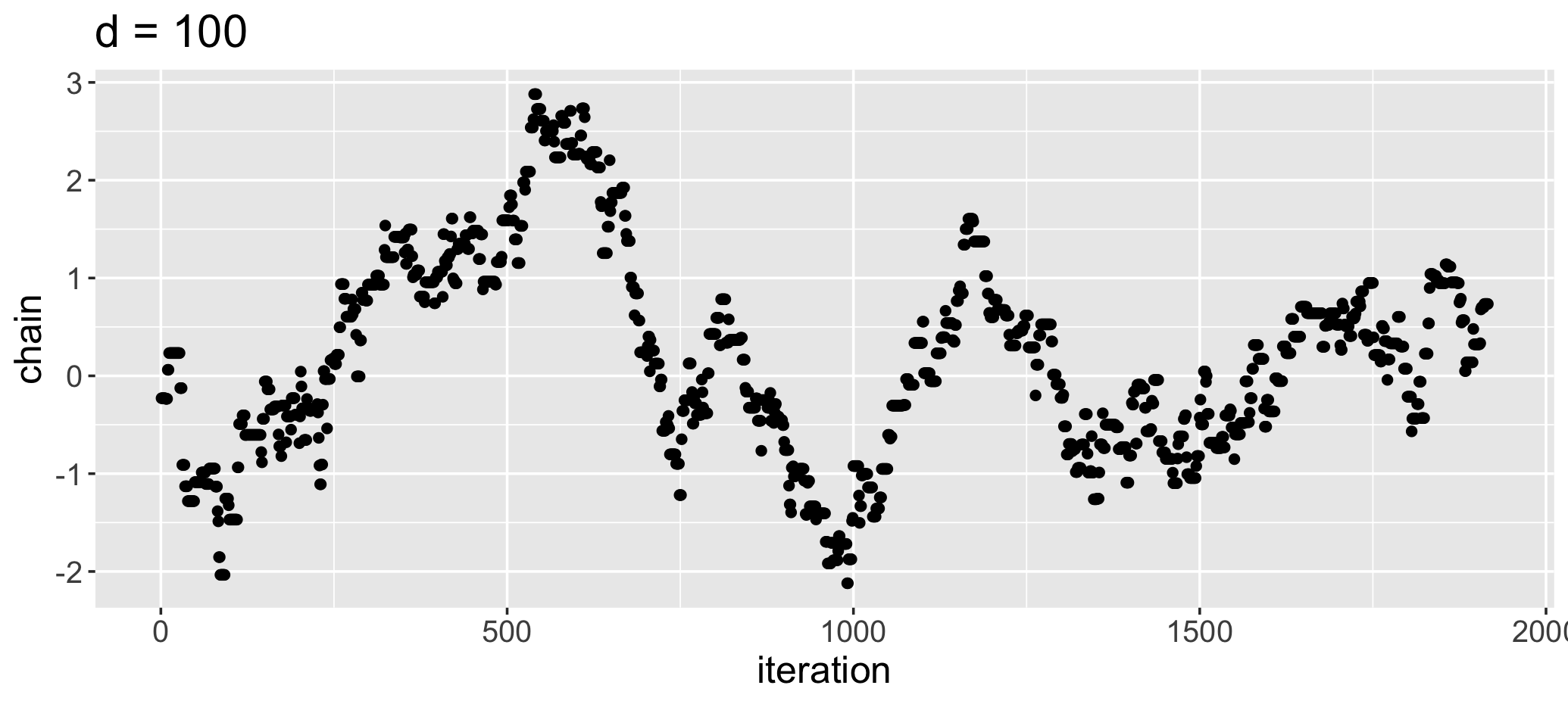}
    \includegraphics[scale = 0.11]{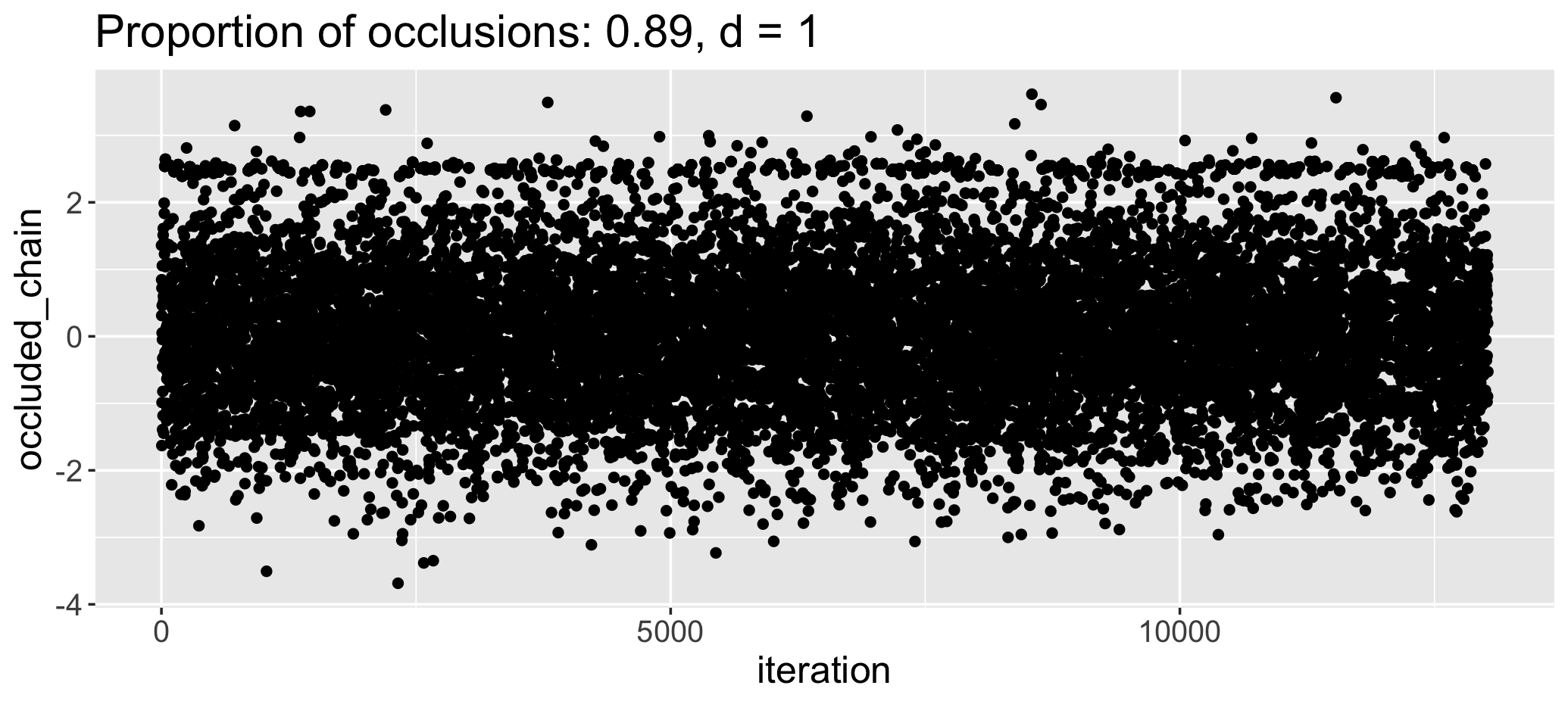}
    \includegraphics[scale = 0.11]{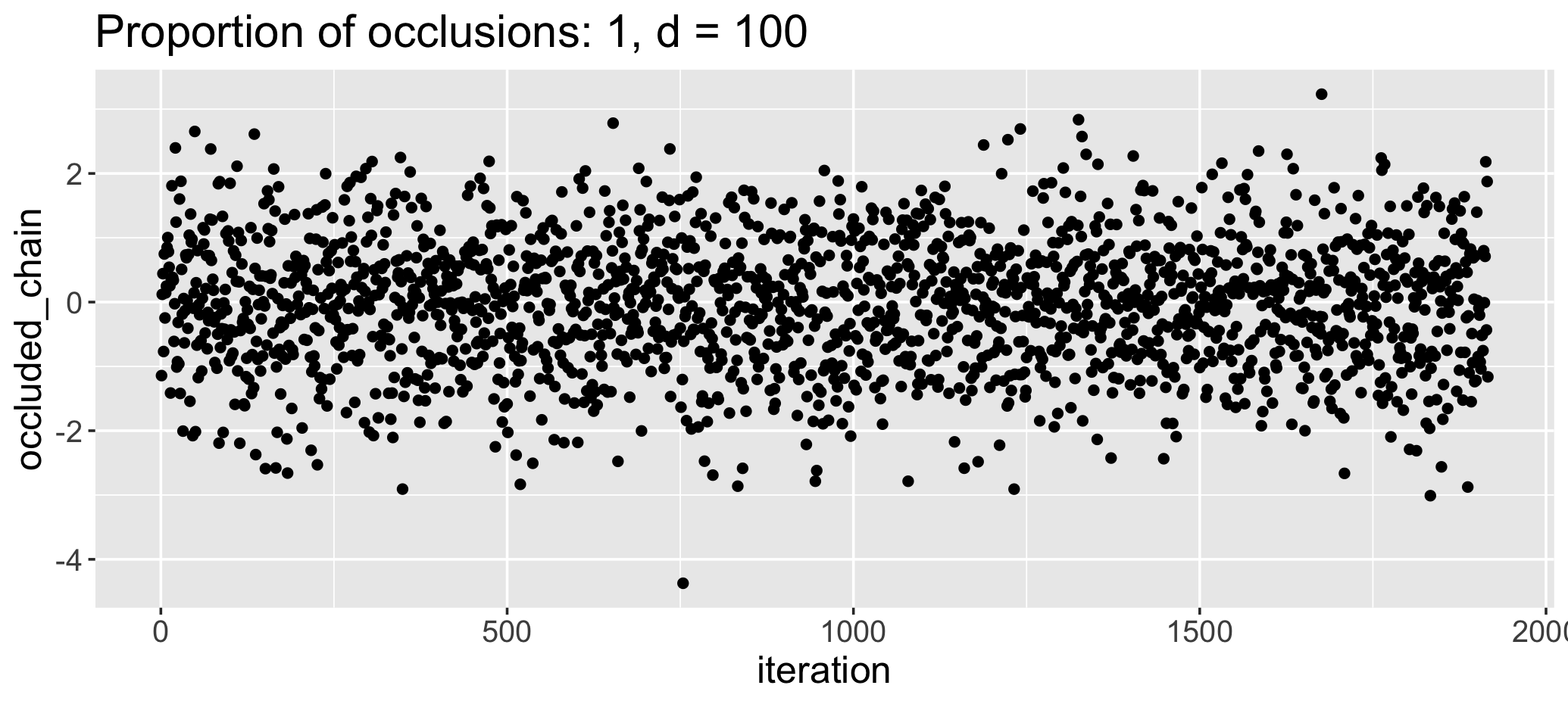}
    \includegraphics[scale = 0.11]{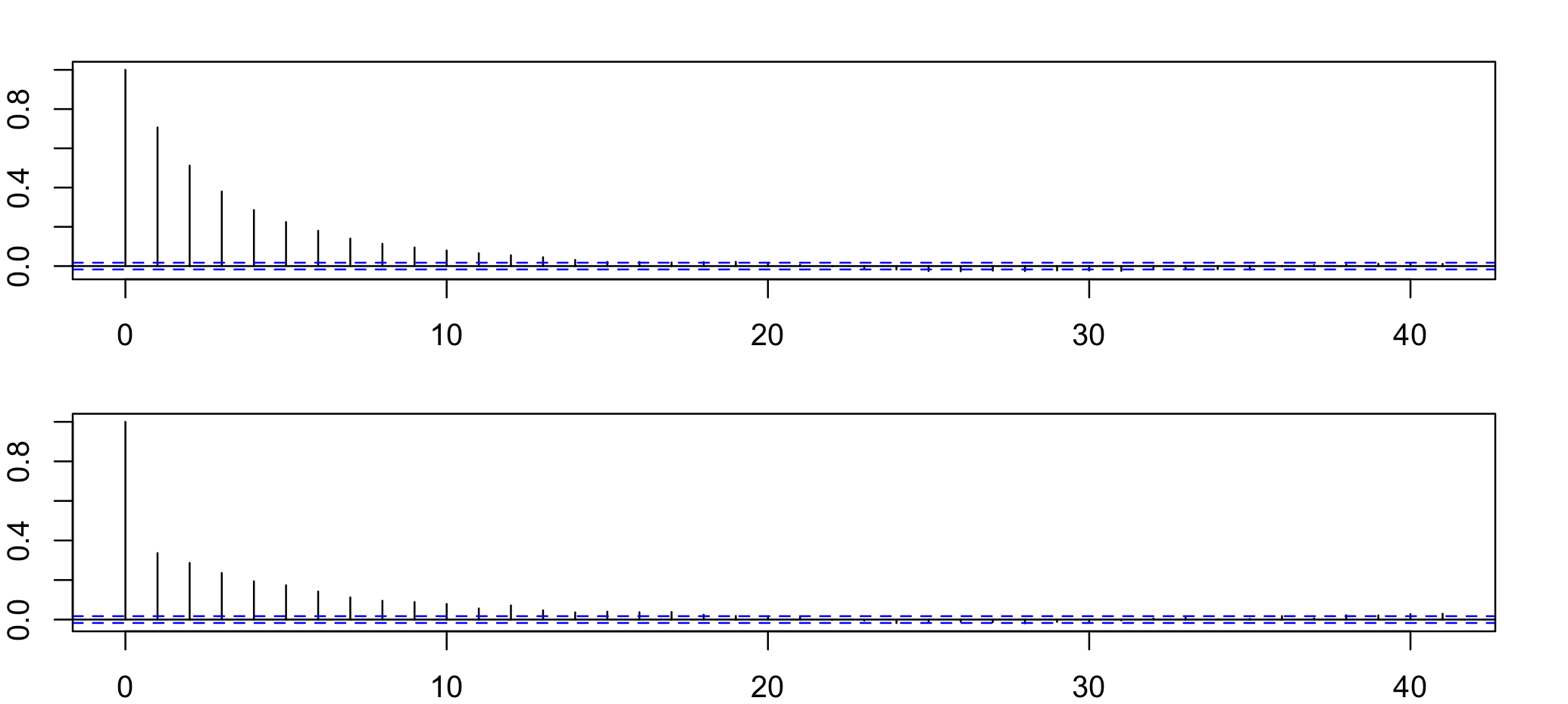}
    \includegraphics[scale = 0.11]{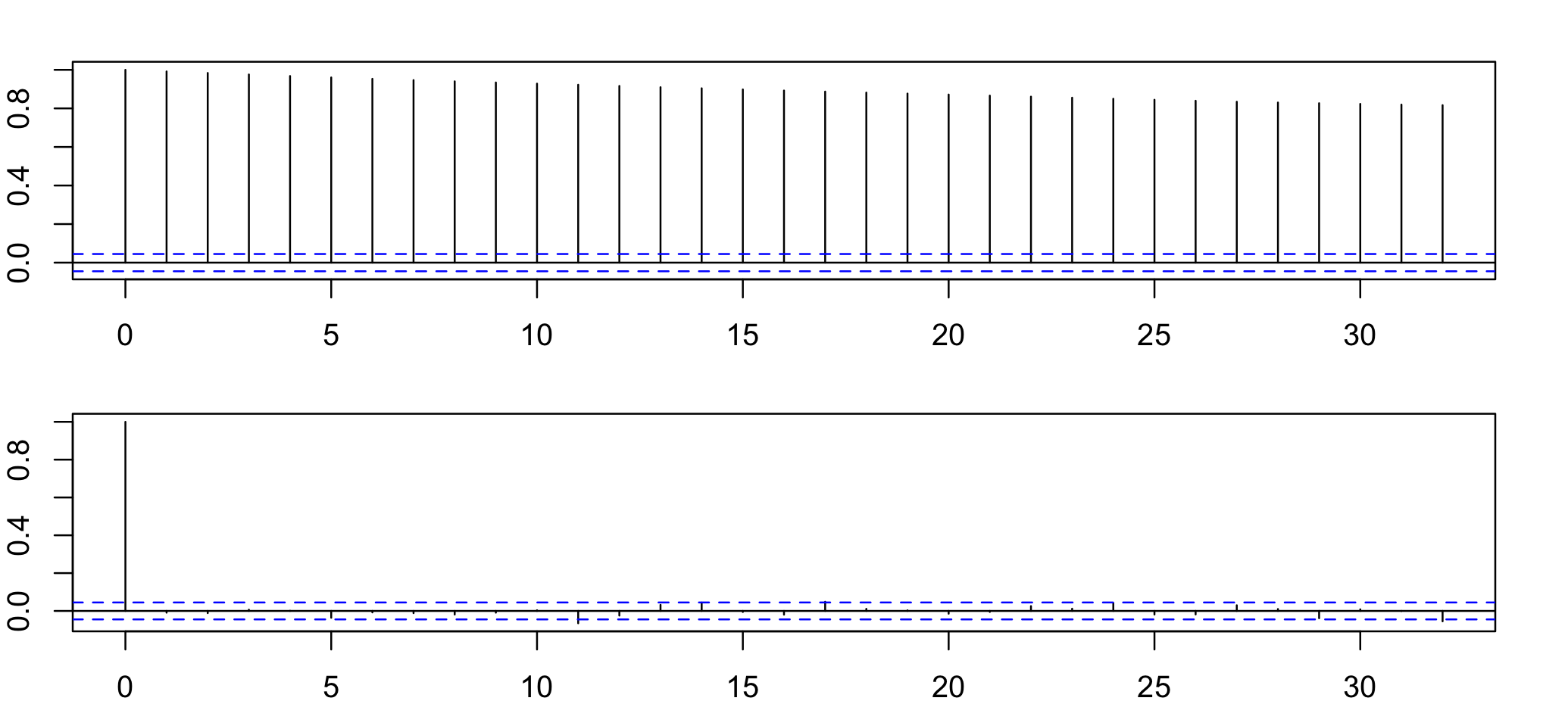}
    \caption{Results from the $d = 1$ case of sampling from (\ref{eqn:bimodal_gaussian_density}) (left) and the $d = 100$ case (right). The top row shows the first component of the RWM chains, the next row shows the first component of $\{\mathds{1}\{S_t = 0\}X_t + \mathds{1}\{S_t = 1\}Y_t\}_{t = 1}^n$, the next row shows autocorrelation function plots of the first component of the RWM chains, the bottom row shows autocorrelation function plots of the first component of $\{\mathds{1}\{S_t = 0\}X_t + \mathds{1}\{S_t = 1\}Y_t\}_{t = 1}^n$.}
    \label{fig:bimodal_gaussian_results}
\end{figure}

In Figure \ref{fig:bimodal_gaussian_results} we show the results of the two experiments, with the $d = 1$ results on the left and the $d = 100$ results on the right. The top row shows the first component of the RWM chains $\{X_t\}_{t = 1}^n$, the row below shows the first component of the occluded chain $\{\mathds{1}\{S_t = 0\}X_t + \mathds{1}\{S_t = 1\}Y_t\}_{t = 1}^n$. In the titles of these plots is the proportion of total states in the chain that were occluded: in the notation of Section \ref{paragraph:general_strategy} this is $(N_1 + \cdots + N_R) / n$.  The bottom four plots show the autocorrelation functions of the two processes. The top row has the autocorrelation function of the first component of the RWM chain, and the bottom row has the autocorrelation function of the first component of the occluded chain.

The main impression given by the figure is that the addition of the occlusions has the effect of decorrelating the process. This is evinced in the traceplots most clearly in the $d = 100$ case, where the RWM chain is clearly displaying the highly autocorrelated behaviour of a diffusion, whereas the states in the occluded process are decorrelated from each other. This is further evinced by the autocorrelation function plots, where in every case at every lag the autocorrelation of the RWM chain is greater than that of the occlusion process. Again this is markedly so in the $d = 100$ case.

We note that although the use of occlusion decorrelates the underlying process, in this case it does not debias it. See, for instance, the $d = 100$ plots on the right hand side, where the RWM chain should spend $10\%$ of its time in the second component of the target whereas the traceplots (and inspection of the output) shows that it spends $0\%$. Therefore the occluded chain also spends no time in the second component, due to the fact that by definition it only ever visits regions that the RWM also visits.

\subsection{Ising Model}\label{subsection:Ising_model}

The Ising model describes the behaviour of magnetic spins $\sigma_i \in \{-1, 1\}$ at nodes in a graph $(V, E)$. Classically the graph in question is a regular cubic lattice although the model has been generalised to arbitrary graphs \citep{delgado:15, bresler:15, mosseri:15}. We assume that no external magnetic field is present, and that the graph is finite with $|V| = N$ vertices. A single state $\sigma\in\{-1,1\}^{N}$ describes the spin configuration
over all vertices, and its potential energy can be written as $U(\sigma)=-\sum_{i=1}^{N}\sum_{j\in S_{i}}J_{ij}\sigma_{i}\sigma_{j}$,
where $S_{i}$ is the set of $i$'s neighbours, and $J_{ij}\in\mathbb{R}$
describes the interaction strength between vertices $i$ and $j$. Here $P$
is the distribution over spin configurations and is defined with
\begin{equation}\label{eqn:Ising_measure}
    P(A):=\sum_{\sigma\in A}\frac{\exp(-\beta U(\sigma))}{Z(\beta)}
\end{equation}
for all subsets $A\subseteq\{-1,1\}^{N}$, where $\beta\in[0,\infty]$
is the inverse temperature and $Z(\beta):=\sum_{\sigma\in\{-1,1\}^{N}}\exp(-\beta U(\sigma))$
is the so called `partition function' i.e. the normalising constant.
We assume that $J_{ij}>0$ for all $i,j\in[N]$ and hence that the system is ferromagnetic.
For such a model neighbouring spins align since doing so decreases
the potential energy of the system. We also assume that $J_{ij} = 1$ for all $(i,j)\in E$. Note that for all $A\subseteq \{-1,1\}^N$ we have that $P(A) = P(-A)$. Therefore any sampler on $P$ should spend an equal amount of time in $A$ and $-A$ in the large sample limit.

If the temperature $\beta^{-1}$ is very low, the strength of alignment between neighbouring spins is very high. Therefore in this regime, if the graph is sufficiently well connected, the sampling algorithm should spend most of its time in configurations where a large proportion of the spins are aligned. It should spend about half the time with the overwhelming majority of the spins aligned in the positive direction, and half of the time with the overwhelming majority of the spins aligned in the negative direction.

If the temperature is very high, the strength of alignment is greatly reduced. Even if the graph is well connected, the proportion of aligned spins should be closer to a half for the running time of the sampler.

A quantity of interest in the Ising model is the \emph{average magnetisation}. It is defined as
\begin{equation}\label{eqn:magnetisation}
    M := \mathbb{E}_P\left[\frac{1}{N}\sum_{i = 1}^N\sigma_i\right]
\end{equation}
and describes the average spin across the entire graph measured in a typical configuration. Since the expression in the expectation is an odd function of $\sigma$ and $P$ is even, we are able to exactly calculate that $M = 0$ for all graphs, at all temperatures. This allows us to objectively measure the output of our sampling algorithms, and of the occlusion process.

\subsubsection{Sampling from the Ising model}

\paragraph{The Metropolis algorithm}

The state space has $2^N$ configurations rendering $Z(\beta)$ very difficult to calculate for large $N$. For this reason practitioners use MCMC methods to sample from the Ising model. One such method is the Metropolis algorithm \citep{metropolis:53}. Originally conceived to sample from the two-dimensional hard-disk model, the Metropolis algorithm can also be applied to the Ising model. It selects a vertex at random and proposes switching the sign of the spin at that vertex. The switch is accepted with the usual Metropolis acceptance probability. For an exact description of the process see Algorithm \ref{alg:metropolis}.

\begin{algorithm}
\caption{Metropolis algorithm applied to the Ising model}\label{alg:metropolis}
\SetKwInOut{Input}{input}\SetKwInOut{Output}{output}
\Input{Chain length $n$, Initial state $\sigma^{(0)} \in \{-1, 1\}^N$, matrix of interaction strengths $J \in \mathbb{R}^{N \times N}$, inverse temperature $\beta \in [0, \infty]$}
\Output{A $P$-invariant Markov chain $\{\sigma^{(t)}\}_{t=1}^n$, where $P$ is as in (\ref{eqn:Ising_measure})}
\For{$t \in [n]$}{
\begin{enumerate}
    \item Sample a vertex $I\sim \textup{Uniform}[N]$.
    \item Let $\sigma^{(t)}_\textup{prop}$ be $\sigma^{(t - 1)}$ with a sign change in the $I$th vertex.
    \item With probability
    \begin{equation*}
        \alpha\left(\sigma^{(t - 1)} \to \sigma^{(t)}_\textup{prop}\right) := \min\left\{1, \exp\left(-\beta U(\sigma^{(t)}_\textup{prop}) + \beta U(\sigma^{(t - 1)})\right)\right\}
    \end{equation*}
    set $\sigma^{(t)} = \sigma^{(t)}_\textup{prop}$ otherwise set $\sigma^{(t)} = \sigma^{(t - 1)}$.
\end{enumerate}
}
\end{algorithm}

Say we apply the Metropolis algorithm to the low temperature setting: when it has reached equilibrium it will likely be in a region of the state space with most of the spins aligned. After this point most of the proposal spin flips are rejected because they are in conflict with the spins of their neighbours. Therefore the chain will spend the vast majority of time in this region of the state space, instead of only half the time. Also any estimate we derive from the chain will have a high variance due to the size of the autocorrelations.

In the high temperature regime the strength of alignment between neighbouring spin sites will only be weak. Therefore the probability of accepting the individual spin flips proposed by the Metropolis algorithm will be higher and the Markov chain generated will exhibit lower autocorrelations than in the low temperature regime.

\paragraph{The Wolff algorithm}

In order to remedy this problem algorithms have been conceived that switch the signs of spins at a cluster of vertices in the graph in a single step of the Markov chain. This is how the Wolff algorithm works \citep{wolff:89}. Specifically we select a vertex at random as the starting vertex in a cluster. We then add its neighbours that have a similar sign to the cluster each with probability $1 - \exp(-2\beta J)$. We add their aligned neighbours (so long as they have not been considered before) in a similar fashion, and their neighbours, etc. etc. until there are no unvisited aligned neighbours to consider. Once we have built a cluster in this way, we deterministically flip all of the spins inside the cluster. In order to formally describe the algorithm in \ref{alg:wolff}, we define the aligned neighbour function $\mathsf{AN}:\mathcal{P}(V) \to \mathcal{P}(V)$ which takes a subset of vertices of all the same sign spin, and outputs all those neighbours of the subset which are aligned.

\begin{algorithm}
\caption{Wolff algorithm}\label{alg:wolff}
\SetKwInOut{Input}{input}\SetKwInOut{Output}{output}
\Input{Chain length $n$, Initial state $\sigma^{(0)} \in \{-1, 1\}^N$, matrix of interaction strengths $J \in \mathbb{R}^{N \times N}$, inverse temperature $\beta \in [0, \infty]$, aligned neighbour function $\mathsf{AN}$}
\Output{A $P$-invariant Markov chain $\{\sigma^{(t)}\}_{t=1}^n$, where $P$ is as in (\ref{eqn:Ising_measure})}
\For{$t \in [n]$}{
\begin{enumerate}
    \item Sample a vertex $I\sim \textup{Uniform}[N]$.
    \item Initialise the cluster $C = \{I\}$, a set of visited vertices $\mathcal{V} = \{I\}$, a set of aligned neighbours which have not been previously visited $\mathcal{N} = \mathsf{AN}(\{I\})\backslash \mathcal{V}$, and a boolean \texttt{neighbours} that is \texttt{true} if $\mathcal{N}$ is nonempty and $\texttt{false}$ otherwise.
\end{enumerate}
\While{\texttt{neighbours}}{
    \For{$j \in \mathcal{N}$}{
    \begin{enumerate}
        \item With probability $1-\exp(-2\beta J)$ append $j$ to $C$.
        \item Append $j$ to $\mathcal{V}$.
    \end{enumerate}
    }
    \begin{enumerate}
        \item Set $\mathcal{N} = \mathsf{AN}(C) \backslash \mathcal{V}$.
        \item If $\mathcal{N} = \varnothing$ set $\texttt{neighbours} = \texttt{false}$.
    \end{enumerate}
}
Set $\sigma^{(t)} = \sigma^{(t - 1)}$ and flip all the signs of the spins at the vertices in the cluster $C$ in $\sigma^{(t)}$. 
}
\end{algorithm}

In an above paragraph we explained why the Metropolis algorithm would work poorly in the low temperature setting. Let's say we use the Wolff algorithm instead. After reaching equilibrium, if the graph is sufficiently connected, most of the spins will be aligned. The value of the $\beta$ parameter will dictate that the cluster inclusion probability $1 - \exp(-2\beta J)$ will be high. The number of aligned visitors will also be high. Therefore the Wolff algorithm will build large clusters whose spins it will flip deterministically, allowing it to traverse large distances in the state space in a single step.

In the high temperature regime the strength of alignment between spins is weaker. Hence in equilibrium the size of aligned clusters will generally be smaller, and so the individual moves generated by the Wolff algorithm will involve a smaller amount of spin flips. Therefore in the high temperature regime we expect the Wolff algorithm to exhibit higher autocorrelation than in the low temperature regime.

\subsubsection{An efficiently simulable variational approximation to the Ising model}\label{subsubsection:an_efficiently_simulable_variational_approximation_to_the_Ising_model}

\begin{figure}[h]
    \centering    \includegraphics[scale = 0.15]{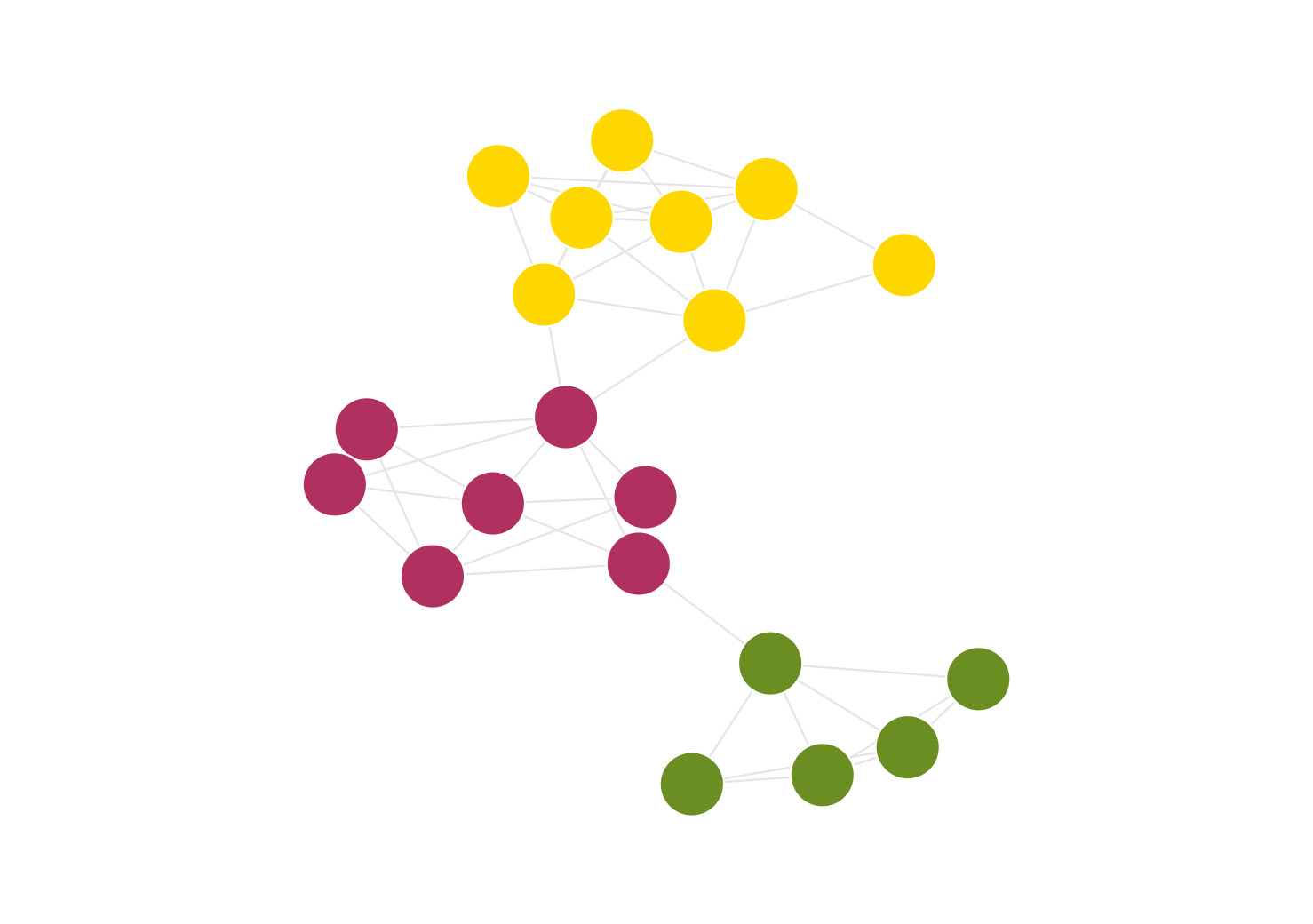}
    \includegraphics[scale = 0.15]{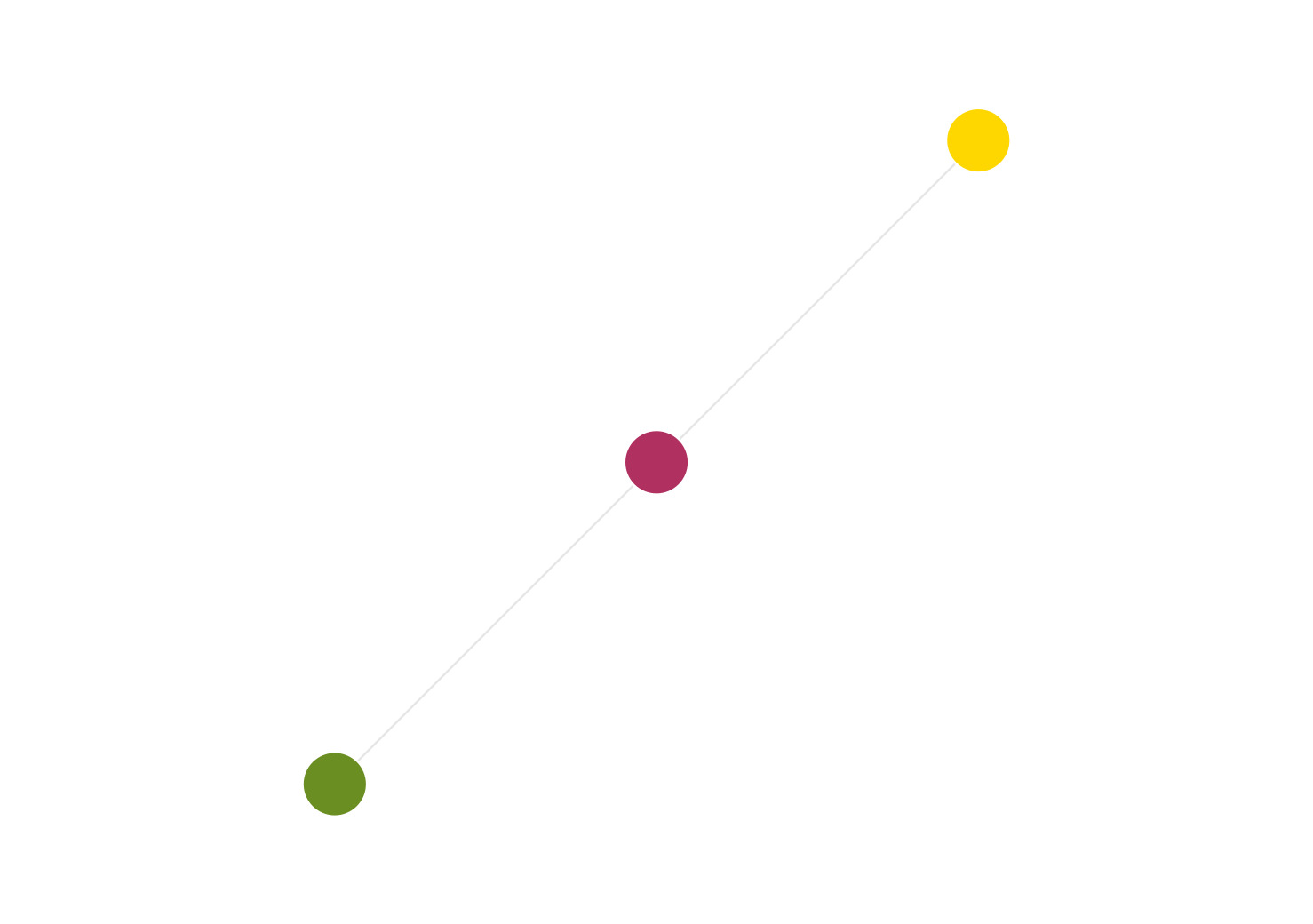}
    \caption{Left: $(V, E)$, right: $(\tilde{V}, \tilde{E})$. The colours correspond to the $V_i$'s in the original $(V, E)$ that get collapsed to nodes in $(\tilde{V}, \tilde{E})$. Where there exist any edges between two $V_i$'s in $(V, E)$, there is an edge between the corresponding nodes in $(\tilde{V}, \tilde{E})$.}
    \label{fig:two_Ising_graphs}
\end{figure}

In order to simulate the occlusion process described in Algorithm \ref{alg:embarrassingly_parallel} we need access to a variational approximation to the Ising model which we can easily sample from. To do this we form a partition of $V=\bigcup_{i=1}^{k}V_{i}$ such that $k$
is small, and define a new graph $(\tilde{V},\tilde{E})$ where $|\tilde{V}|=k$
and $(i',j')\in\tilde{E}$ when there exists an edge in the original
graph $(i,j)\in E$ such that $i\in V_{i'}$ and $j\in V_{j'}$. The
node clusters $V_{i}$ therefore form the nodes in $\tilde{V}$. See Figure \ref{fig:two_Ising_graphs} for an example of two graphs: $(V, E)$ and $(\tilde{V}, \tilde{E})$.

We
extend the state space to include $\mu\in\{-(1-\epsilon),1-\epsilon\}^{k}$ for a given $\epsilon > 0$
such that each $\mu_{i}$ serves as the mean of the spins in $V_{i}$.
We let $\mu$ behave according to the dynamics of an Ising model on
$(\tilde{V},\tilde{E})$ and sample the spins within the clusters
$V_{i}$ independently with mean $\mu_{i}$. Therefore we dictate
that
\[
\tilde{U}(\mu):=-\sum_{i=1}^{k}\sum_{j:(i,j)\in\tilde{E}}\tilde{J}_{ij}\mu_{i}\mu_{j}
\]
where each $\tilde{J}_{ij}\in\mathbb{R}$ possibly depends on the
number of edges between subgraphs $i$ and $j$. Since $k$ is small
we can calculate the normalising constant 
\[
\tilde{Z}(\tilde{\beta}):=\sum_{\mu\in\{-(1-\epsilon),1-\epsilon\}^{k}}\exp(-\tilde{\beta}\tilde{U}(\mu))
\]
 where $\tilde{\beta}\in[0,\infty]$ is an inverse temperature hyperparameter.
As mentioned before, we sample the spins within each subgraph independently
conditional on their means, which are given by $\mu$. We choose a
Rademacher distribution such that
\[
q(\sigma\left|\mu\right.):=\prod_{i=1}^{k}\prod_{s=1}^{|V_{i}|}\left(\frac{1+\mu_{i}}{2}\right)^{\frac{1+\sigma_{s}^{(i)}}{2}}\left(\frac{1-\mu_{i}}{2}\right)^{\frac{1-\sigma_{s}^{(i)}}{2}}
\]
where $\sigma_{s}^{(i)}\in\{-1,1\}$ is the $s$th spin in the $i$th
cluster. Therefore the full variational distribution has density
\[
q(\sigma):=\sum_{\mu\in\{-(1-\epsilon),1-\epsilon\}^{k}}q(\sigma\left|\mu\right.)\tilde{Z}(\tilde{\beta})^{-1}\exp(-\tilde{\beta}\tilde{U}(\mu))
\]
To sample from $Q$ we simply sample $\mu$ according to the Ising distribution
defined by $\tilde{U}$ and then sample $\sigma$ conditionally on
$\mu$. To do this we need to take an exact sample of $\mu$. For
this we can simply calculate the probabilities of all points in $\{-(1-\epsilon),1-\epsilon\}^{k}$
since we have access to $\tilde{Z}(\tilde{\beta})$.

\subsubsection{Experiment setup}

In order to test the occlusion process in a full range of settings, we record its performance on the Ising model at low and high temperatures on graphs of different sizes and connectivities, and we compare it with the Metropolis and the Wolff algorithm.

To generate the underlying graph we sample from an $N$-vertex stochastic block model \citep{lee:2019} with $k$ communities, where the community sizes are sampled uniformly from size $k$ partitions of $\{1,...,N\}$. For the matrix of edge probabilities $A \in \mathbb{R}^{k \times k}$ we have $A_{ii} = 0.8$ for all $i \in [k]$ and $A_{ij} = 0.01$ for all $i \neq j$ and $i,j \in [k]$. The left side of Figure \ref{fig:two_Ising_graphs} shows such a graph with $k = 3$ communities and $N = 20$ vertices.

As stated in the introduction of the model, we set $J_{ij} = J = 1$ for all $i,j\in[n]$. We set $\beta = 1$ to emulate a low temperature setting, and $\beta = 0.01$ to emulate a high temperature setting. For the variational approximation defined in Section \ref{subsubsection:an_efficiently_simulable_variational_approximation_to_the_Ising_model} we define the clusters $V_i$ as the communities of the underlying graph. In practice we will not have access to the actual latent cluster structure of the graph, and therefore some clustering algorithm must be used to find these. We set $\tilde{J}_{ij}$ to the number of connections between clusters $V_i$ and $V_j$ for all $i, j \in [k]$ and $\tilde{\beta} = 0.5 \times \beta$ in all temperature settings. We set $\epsilon = 0.1$ in the $\beta = 1$ low temperature setting and $\epsilon = 0.9$ in the $\beta = 0.01$ high temperature setting.

For the number of regions we set $R = 3$. To choose the constants $C_1 < C_2$ which define the regions (see Section \ref{subsubsection:sampling_from_the_P_is}) we run the Wolff algorithm for 20 seconds from a random $\sigma^{(0)} \sim \textup{Uniform}\{-1,1\}^N$ initialisation. We set $C_1$ to be the median from $\{dP/dQ(\sigma^{(t)})\}_{t \geq 1}$ and $C_2$ to be the maximum.

To initialise the Markov chains generated by the Metropolis and Wolff algorithms we use the final state in a Markov chain generated by the Swendsen-Wang algorithm. In \cite[Theorems 1 and 2]{huber:1999} one can find conditions under which the Swendsen-Wang chain couples with probability $\geq 1/2$ for a specified chain length. We therefore use these results to run a Swendsen-Wang chain until the probability of being uncoupled is $\leq 0.0001$ for each model we choose to sample from.

For each pair $(k, N)$ with $k \in \{2, 5, 10\}$ and $N \in \{20, 50, 100\}$ we run 15 replications of the Metropolis and Wolff algorithms for 20 seconds each. Each time we use one of these algorithms, we also run the occlusion process defined in Algorithm \ref{alg:embarrassingly_parallel} so that we can compare their results. Specifically, we compare their estimates of the expected magnetisation as defined in (\ref{eqn:magnetisation}).

\subsubsection{Experiment results}\label{subsubsection:Ising_experiment_results}

\begin{figure}
    \centering
    \includegraphics[scale = 0.9]{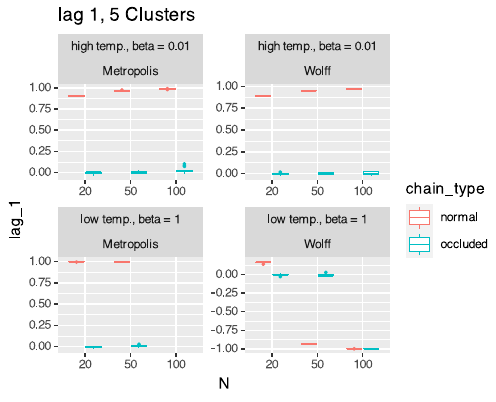}
    \includegraphics[scale = 0.9]{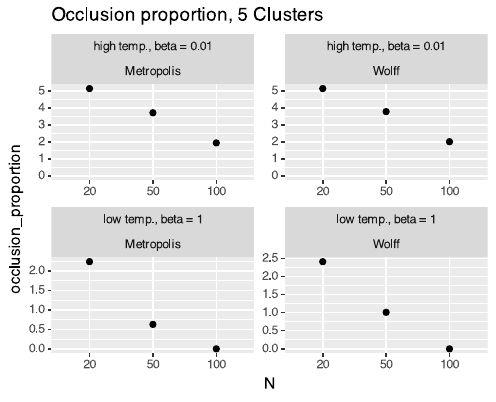}
    \includegraphics[scale = 1.3]{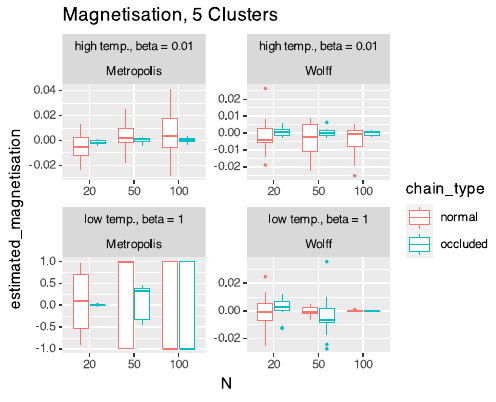}
    \caption{Three graphs comparing the performance of the occlusion process with the Metropolis and Wolff algorithms on the Ising model at a variety of temperatures, for a variety of graph sizes. In every case the horizontal axes show the number of vertices $N$ in the graphs. Bottom: the vertical axes denote the algorithm's estimates of the expected magnetisation. Top left: the vertical axes denote the lag 1 autocorrelation coefficient of the magnetisation over the states produced by the algorithms. Top right: the vertical axes show the number of samples from the $P_i$'s in Algorithm \ref{alg:embarrassingly_parallel} divided by the number of states in the Markov chain $n$. We magnify the estimated magnetisation plot for ease of comprehension, the top two plots then help to explain the phenomena in the bottom plot.}
    \label{fig:Ising_results}
\end{figure}

In Figure \ref{fig:Ising_results} we show three graphs all from the same set of results which are generated according to the setup described above. Here we display the results in the $k = 5$ case. For the $k \in \{2, 10\}$ cases, refer to Appendix \ref{appendix:additional_results_for_the_Ising_experiment_in_section}. Each graph has four subgraphs across the various temperatures and algorithms we use to sample. The `normal' in the legend refers to data from the non-occluded Markov chain. The bottom graph compares the estimates of the expected magnetisation between the occluded process and the Metropolis and Wolff algorithms. The true value of the expected magnetisation is 0. The top left graph shows the lag 1 autocorrelation coefficient of the magnetisation over the course of the Markov chains. The top right graph shows the number of occluded states as a proportion of the total number of states from the Markov chain sampling algorithms. In the notation of Section \ref{paragraph:general_strategy} this is $(N_1 + \cdots + N_R) / n$.

As expected the Wolff algorithm excels in the low temperature setting whereas the Metropolis algorithm fares very poorly due to the fact that in each instance it never strays from its initial position. In the high temperature setting the variance of the estimates from the Wolff algorithm are lower in general than those from the Metropolis chain, but the difference is not as stark as in the low temperature setting. An explanation as to why the Wolff algorithm does so well in the low temperature setting is offered by the graphs of the lag 1 autocorrelation coefficients (top right). One can clearly see that some of these coefficients are close to -1, explaining why the variance of the resulting estimates is very low.

As concerns the performance of the occluded chain versus the Metropolis and Wolff samplers: we can see that in the high temperature setting, in every case the variance of the occluded estimator is much lower than those of the Markov chains. The fact that the lag 1 autocorrelation coefficients of the occluded process are so close to zero, combined with the high proportions of occlusion events as seen in the top level of the top right, explain this fact.

In the low temperature setting the story is more mixed. If we compare the performance of the occluded estimator with the Metropolis algorithm, we can see that for $N = 20$ and $50$ the occluded estimator offers a reduction in variance. This is explained by the corresponding reduction in the lag 1 autocorrelation coefficients and the high occlusion proportions. However the low occlusion proportion in the $N = 100$ case means that the occluded chain and the Metropolis chain are basically the same, and so the occluded estimator offers no increase in performance. In this case the Markov chain does not move, and so the lag 1 estimator breaks, as seen in the graph in the top left.

Comparing the occluded estimator with the Wolff algorithm in the low temperature regime shows that in the $N = 50$ case using the occluded estimator actually decreases the performance by increasing the bias and the variance. This is explained by the fact that the occlusion events tend to \emph{de}correlate subsequent states in the process, whereas the Wolff algorithm produces \emph{anti}correlated states. Hence the decorrelations are undesirable in this case. Again, as with the Metropolis algorithm, the number of occlusions in the $N = 100$ case is very few, and there is negligible difference between the performance of the occlusion process and the Wolff algorithm.

We would like to note that we have not tuned the parameters of the variational distribution for any of the models presented here. Therefore for a better tuned variational distribution, we would expect the occlusion proportions to be higher, and so at least in the low temperature $N = 100$ case achieve a reduced lower variance of the occluded estimator compared to the Metropolis algorithm.

\subsubsection{Satisfaction of theoretical conditions for variance reduction}\label{subsubsection:satisfaction_of_theoretical_conditions_for_variance_reduction}

Here we claim that, in the high temperature setting, we satisfy condition 2. from Proposition \ref{prop:ideal_variance_dominance}, such that the variance reductions seen in Figure \ref{fig:Ising_results} verify the theory.

Clearly the measure defining the Ising model (\ref{eqn:Ising_measure}) is even in its arguments and the magnetisation (\ref{eqn:magnetisation}) is odd. That the regions satisfy $\mathsf{X}_i = -\mathsf{X}_i$ follow from the fact that the Radon-Nikodym derivative is such that $dP/dQ(\sigma) = dP/dQ(-\sigma)$ for all $\sigma \in \{-1,1\}^N$. These three facts satisfy the conditions outlined in Example \ref{ex:odd_function_even_measure} such that $f \equiv \overleftarrow{P}f + P(f)$. That the Markov kernel is positive when applied to the magnetisation functional is evidenced by the lag-1 autocorrelation coefficients in the high temperature settings of Figure \ref{fig:Ising_results}. The last criterion for variance reduction we need is for $\hat{\mu}_\textup{occ} = \hat{\mu}_\textup{ideal}$. This is evidenced by the occlusion proportion graphs in Figure \ref{fig:Ising_results}, which imply that $\alpha \equiv 1$ at high temperature.

Overall this suggests that in the high-temperature setting, condition 2. of Proposition \ref{prop:ideal_variance_dominance} is satisfied, and so the theory suggests that $\textup{Var}(\hat{\mu}_{\textup{occ}}) \leq \textup{Var}(n^{-1}\sum_{t=1}^{n}f(X_{t}))$. This is verified by the variance reductions seen at high-temperature in Figure \ref{fig:Ising_results}.

\section{Discussion}

\subsection{Summary}

In this paper we have introduced the occlusion process which sits on top of an existing Markov kernel $K$. It produces unbiased estimates of expectations under the equilibrium distribution $P$ of $K$. The occlusion process is designed to reduce the variance of the estimates produced solely by $K$ at no additional compute-time cost. It uses a variational distribution $Q$ to do this.

We define the process in the Markov kernel (\ref{eqn:K_occ_definition}), showing that it is unbiased and stating its variance in Proposition \ref{prop:variance_mu_occ}. In Section \ref{section:inherited_theoretical_properties_of_the_occlusion_process} we show that it inherits properties such as an LLN (Theorem \ref{thm:LLN_inheritance}), convergence in a normed function space (Theorem \ref{thm:inherited_normed_function_convergence}), and geometric ergodicity (Theorem \ref{thm:inherited_geometric_ergodicity}) from $K$. In Section \ref{section:efficient_simulation_of_the_occlusion_process} we explain how to simulate the occlusion process at no additional computational cost in terms of wall-clock time. In Section \ref{section:numerical_experiments} we present two numerical experiments so as to inspect the empirical properties of the occlusion process, and to compare its performance against the Markov chains generated by $K$. The first experiment in Section \ref{subsection:bimodal_gaussian_mixture} compares the ability of $K$ with the occlusion process to sample from a bimodal Gaussian mixture, for which we have a good approximation $Q$ of one of the mixture components. The results show that the occlusion process is able to effectively decorrelate subsequent steps in the Markov chain generated by $K$. In Section \ref{subsection:Ising_model} we define the Ising model on an arbitrary graph, and introduce the Metropolis (Algorithm \ref{alg:metropolis}) and Wolff (Algorithm \ref{alg:wolff}) algorithms which are used to form Markov chains with which we form estimators for this model. Since the occlusion process uses a variational distribution $Q$ of the Ising model we propose one in Section \ref{subsubsection:an_efficiently_simulable_variational_approximation_to_the_Ising_model}. We run the Metropolis and Wolff at various temperatures on various graphs, and show that the occlusion process produces estimators of reduced variance, apart from in the low temperature case with the Wolff algorithm. This is because, in such a setting the Wolff algorithm produces states which are anticorrelated, whereas the occlusion process produces states which are decorrelated. In Section \ref{subsubsection:satisfaction_of_theoretical_conditions_for_variance_reduction} we argue that in the high temperature setting, we satisfy the conditions for variance reduction, which explains the variance dominance in the experimental results.

\subsection{Extensions}

There are numerous interesting extensions to the work presented here. On the theoretical side, it would be interesting if we could establish conditions under which the variance of the occluded estimator \ref{eqn:mu_hat_occ} is dominated by the variance of the estimator produced by the Markov kernel $K$ (where $K$ is positive). In the numerical experiments we clearly see variance dominance in multiple cases, so it would be useful to theoretically establish this phenomenon. In general it would be interesting to find the minimum variance unbiased estimator of an expectation $\mu$ under $P$ given an array of samples $\{X_t\}_{t = 1}^n$ from a $P$-invariant Markov chain and $R$ collections of samples $\{\{Y_{ij}\}_{j = 1}^{N_i}\}_{i = 1}^R$ with $Y_{ij} \sim P_i$ for all $j \in [N_i]$ and $i \in [R]$. For example: in the embarrassingly parallel process outlined in Algorithm \ref{alg:embarrassingly_parallel} the $Y_{ij}$ samples occlude uniformly at random the appropriate $X_t$ samples. Is there some allocation scheme that is less random, achieves a lower variance, but is still provable unbiased? We might also not want to throw away the samples $X_t$ from the Markov chain upon occlusion. Is there some estimator then that uses all the samples from the Markov chain, as well as the $Y_{ij}$ samples, in an unbiased way?

On the practical side of things there is a plethora of ways in which we could enhance the occlusion process. One way is to learn the optimal tuning parameters of the variational distribution $Q$ online as the process runs. Another way is to use the samples from the $P_i$'s to inform the Markov chain generated by $K$ as in the equi-energy sampler \citep{kou:2006}. We could also use enhanced versions of the rejection samplers in Algorithm \ref{alg:embarrassingly_parallel} such as the squeeze method \citep{marsaglia:1977}. 

\section*{Acknowledgements and Disclosure of Funding}

We would like to that Francois Perron and Charly Andral for their discussions. MH is funded by an EPSRC DTP and part of this work was produced as a result of the UKRI UK-Canada Globalink doctoral exchange scheme. FM is funded by NSERC of Canada.

\section{Appendix A: Proofs}

\subsection{Proofs for Section \ref{section:the_occlusion_process}}

\subsubsection{Proof of Proposition \ref{prop:ideal_estimator}\label{proof:ideal_estimator}}

To prove the unbiasedness we have that
\begin{align*}
\mathbb{E}[\hat{\mu}_{\textup{ideal}}] & =\frac{1}{n}\mathbb{E}\left[\sum_{i=1}^{R}\mathbb{E}\left[\sum_{j=1}^{N_{i}}f(Y_{ij})\left|\{N_{i}\}_{i=1}^{R}\right.\right]\right]\\
 & =\frac{1}{n}\mathbb{E}\left[\sum_{i=1}^{R}N_{i}\mu_{i}\right]\\
 & =\frac{1}{n}\sum_{i=1}^{R}\mathbb{E}[N_{i}]\mu_{i}=\frac{1}{n}\sum_{i=1}^{R}nP(\mathsf{X}_{i})\mu_{i}
\end{align*}
and the result follows. For the variance we use the law of total variance:

\[
\textup{Var}(\hat{\mu}_{\textup{ideal}})=\mathbb{E}\left[\textup{Var}\left(\hat{\mu}_{\textup{ideal}}\left|\{\rho(X_{t})\}_{t=1}^{n}\right.\right)\right]+\textup{Var}\left(\mathbb{E}\left[\hat{\mu}_{\textup{ideal}}\left|\{\rho(X_{t})\}_{t=1}^{n}\right.\right]\right)
\]
where $\rho(X_{t})$ is simply the region that $X_{t}$ is in. Inspecting
the second term on the right hand side:
\[
\mathbb{E}\left[\hat{\mu}_{\textup{ideal}}\left|\{\rho(X_{t})\}_{t=1}^{n}\right.\right]=\frac{1}{n}\sum_{i=1}^{R}\sum_{j=1}^{N_{i}}\mathbb{E}[f(Y_{ij})\left|\{\rho(X_{t})\}_{t=1}^{n}\right.]=\frac{1}{n}\sum_{i=1}^{R}\mu_{i}N_{i}
\]
and so
\begin{align*}
\textup{Var}\left(\mathbb{E}\left[\hat{\mu}_{\textup{ideal}}\left|\{\rho(X_{t})\}_{t=1}^{n}\right.\right]\right) & =\textup{Var}\left(\frac{1}{n}\sum_{i=1}^{R}\mu_{i}N_{i}\right)\\
 & =\frac{1}{n^{2}}\sum_{i,i'}^{R}\mu_{i}\mu_{i'}\textup{Cov}\left(N_{i},N_{i'}\right)\\
 & =\frac{1}{n^{2}}\sum_{i,i'}^{R}\mu_{i}\mu_{i'}\sum_{t,t'}^{n}\textup{Cov}\left(\mathds{1}\{X_{t}\in\mathsf{X}_{i}\},\mathds{1}\{X_{t'}\in\mathsf{X}_{i'}\}\right)
\end{align*}
We have that
\[
\sum_{t,t'}^{n}\textup{Cov}\left(\mathds{1}\{X_{t}\in\mathsf{X}_{i}\},\mathds{1}\{X_{t'}\in\mathsf{X}_{i'}\}\right)=n\textup{Cov}_{P}\left(\mathds{1}\{X\in\mathsf{X}_{i}\},\mathds{1}\{X\in\mathsf{X}_{i'}\}\right)+2\sum_{k=1}^{n-1}(n-k)\textup{Cov}_{P}\left(K^{k}\mathds{1}\{X\in\mathsf{X}_{i}\},\mathds{1}\{X\in\mathsf{X}_{i'}\}\right)
\]
where $K$ is the Markov operator of $\{X_{t}\}_{t=1}^{n}$. Since
$\textup{Cov}_{P}\left(\mathds{1}\{X\in\mathsf{X}_{i}\},\mathds{1}\{X\in\mathsf{X}_{i'}\}\right)=\mathds{1}\{i=i'\}P(\mathsf{X}_{i})-P(\mathsf{X}_{i})P(\mathsf{X}_{j})$
we have that
\[
\frac{1}{n^{2}}\sum_{i,i'}^{R}\mu_{i}\mu_{i'}\textup{Cov}_{P}\left(\mathds{1}\{X\in\mathsf{X}_{i}\},\mathds{1}\{X\in\mathsf{X}_{i'}\}\right)=\frac{1}{n}\left(\sum_{i=1}^{R}P(\mathsf{X}_{i})\mu_{i}^{2}-\mu^{2}\right)=\frac{1}{n}\textup{Var}_{P}(\overrightarrow{P}f)
\]
which gives that
\begin{align*}
\textup{Var}\left(\mathbb{E}\left[\hat{\mu}_{\textup{ideal}}\left|\{\rho(X_{t})\}_{t=1}^{n}\right.\right]\right) & =\frac{1}{n}\textup{Var}_{P}(\overrightarrow{P}f)+\frac{1}{n}2\sum_{k=1}^{n-1}\frac{n-k}{n}\textup{Cov}_{P}\left(K^{k}\overrightarrow{P}f(X),\overrightarrow{P}f(X)\right)\\
 & =\textup{Var}\left(\frac{1}{n}\sum_{t=1}^{n}\overrightarrow{P}f(X_{t})\right)
\end{align*}
after we absorb the sums involving $i$ and $i'$ into $\textup{Cov}_{P}\left(K^{k}\mathds{1}\{X\in\mathsf{X}_{i}\},\mathds{1}\{X\in\mathsf{X}_{i'}\}\right)$.
Finally we inspect the first term on the right hand side:
\begin{align*}
\mathbb{E}\left[\textup{Var}\left(\hat{\mu}_{\textup{ideal}}\left|\{\rho(X_{t})\}_{t=1}^{n}\right.\right)\right] & =\mathbb{E}\left[\frac{1}{n^{2}}\sum_{i=1}^{R}\sum_{j=1}^{N_{i}}\textup{Var}\left(f(Y_{ij})\left|\{\rho(X_{t})\}_{t=1}^{n}\right.\right)\right]\\
 & =\mathbb{E}\left[\frac{1}{n^{2}}\sum_{i=1}^{R}N_{i}\sigma_{i}^{2}\right]\\
 & =\frac{1}{n^{2}}\sum_{i=1}^{R}\mathbb{E}[N_{i}]\sigma_{i}^{2}=\frac{1}{n}\textup{Var}_{P}(\overleftarrow{P}f(X))
\end{align*}
from which the full result follows.

\subsubsection{Proof of Lemma \ref{lem:K_variance_dominance_over_resolution}\label{proof:K_variance_dominance_over_resolution}}

First note that $\mathbb{E}[f(X_{t})\left|\{\rho(X_{t})\}_{t=1}^{n}\right.]=\mu_{\rho(X_{t})}=\overrightarrow{P}f(X_{t})$.
Second, by the law of total variance we have that 
\begin{align*}
\textup{Var}\left(\frac{1}{n}\sum_{t=1}^{n}f(X_{t})\right) & =\mathbb{E}\left[\textup{Var}\left(\frac{1}{n}\sum_{t=1}^{n}f(X_{t})\left|\{\rho(X_{t})\}_{t=1}^{n}\right.\right)\right]+\textup{Var}\left(\mathbb{E}\left[\frac{1}{n}\sum_{t=1}^{n}f(X_{t})\left|\{\rho(X_{t})\}_{t=1}^{n}\right.\right]\right)\\
 & =\mathbb{E}\left[\textup{Var}\left(\frac{1}{n}\sum_{t=1}^{n}f(X_{t})\left|\{\rho(X_{t})\}_{t=1}^{n}\right.\right)\right]+\textup{Var}\left(\frac{1}{n}\sum_{t=1}^{n}\mathbb{E}[f(X_{t})\left|\{\rho(X_{t})\}_{t=1}^{n}\right.]\right)\\
 & =\mathbb{E}\left[\textup{Var}\left(\frac{1}{n}\sum_{t=1}^{n}f(X_{t})\left|\{\rho(X_{t})\}_{t=1}^{n}\right.\right)\right]+\textup{Var}\left(\frac{1}{n}\sum_{t=1}^{n}\overrightarrow{P}f(X_{t})\right)
\end{align*}
and the result follows.

\subsubsection{Proof of Proposition \ref{prop:ideal_variance_dominance}\label{proof:ideal_variance_dominance}}

That the condition 1. entails $\textup{Var}(\hat{\mu}_{\textup{ideal}})\leq\textup{Var}(n^{-1}\sum_{t=1}^{n}f(X_{t}))$ follows from Proposition \ref{prop:ideal_estimator}, Lemma \ref{lem:K_variance_dominance_over_resolution} and the fact that $f\equiv\overrightarrow{P}f$ implies $\overleftarrow{P}f\equiv 0$.

For condition 2. the fact that $f \equiv \overleftarrow{P}f + P(f)$ means that $\overrightarrow{P}f \equiv P(f)$. Therefore $\textup{Var}(\hat{\mu}_{\textup{ideal}}) = n^{-1}\textup{Var}_P(\overleftarrow{P}f)$. Compare this with $\textup{Var}(n^{-1}\sum_{t=1}^{n}f(X_{t}))$:
\[
\textup{Var}\left(n^{-1}\sum_{t=1}^{n}f(X_{t})\right) := \frac{1}{n}\textup{Var}_P(\overleftarrow{P}f) + \frac{1}{n}2\sum_{k = 1}^{n - 1}\frac{n - k}{n}\textup{Cov}_P(\overleftarrow{P}f(X), K^k\overleftarrow{P}f(X))
\]
which clearly dominates $\textup{Var}(\hat{\mu}_{\textup{ideal}}) = n^{-1}\textup{Var}_P(\overleftarrow{P}f)$ when $K$ is positive.

\subsubsection{Proof of Proposition \ref{prop:variance_mu_occ}\label{proof:variance_mu_occ}}

For the unbiasedness of $\hat{\mu}_{\textup{occ}}$ we have the following:
\[
\mathbb{E}[\hat{\mu}_{\textup{occ}}]=\frac{1}{n}\sum_{t=1}^{n}\mathbb{E}[\mathds{1}\{S_{t}=0\}f(X_{t})]+\mathbb{E}[\mathds{1}\{S_{t}=1\}f(Y_{t})]
\]
Inspecting the first term in the summand:
\begin{align*}
\mathbb{E}[\mathds{1}\{S_{t}=0\}f(X_{t})] & =\mathbb{E}[\mathbb{E}[\mathds{1}\{S_{t}=0\}f(X_{t})\left|\{X_{t}\}_{t=1}^{n}\right.]]\\
 & =\mathbb{E}[f(X_{t})\mathbb{E}[\mathds{1}\{S_{t}=0\}\left|\{X_{t}\}_{t=1}^{n}\right.]]\\
 & =\mathbb{E}[(1-\alpha(\rho(X_{t})))f(X_{t})]\\
 & =\sum_{i=1}^{R}P(\mathsf{X}_{i})(1-\alpha(i))P_{i}(f)
\end{align*}
Inspecting the second term:
\begin{align*}
\mathbb{E}[\mathds{1}\{S_{t}=1\}f(Y_{t})] & =\mathbb{E}[\mathbb{E}[\mathds{1}\{S_{t}=1\}f(Y_{t})\left|\{X_{t}\}_{t=1}^{n}\right.]]\\
 & =\mathbb{E}[f(Y_{t})\mathbb{E}[\mathds{1}\{S_{t}=1\}\left|\{X_{t}\}_{t=1}^{n}\right.]]\\
 & =\mathbb{E}[\alpha(\rho(X_{t}))f(Y_{t})]\\
 & =\sum_{i=1}^{R}P(\mathsf{X}_{i})\alpha(i)P_{i}(f)
\end{align*}
Incorporating the summands into the sum gives the desired answer.

As for the variance, we make the following decomposition:
\[
\textup{Var}(\hat{\mu}_{\textup{occ}})=\mathbb{E}\left[\textup{Var}\left(\hat{\mu}_{\textup{occ}}\left|\{\rho(X_{t})\}_{t=1}^{n}\right.\right)\right]+\textup{Var}\left(\mathbb{E}\left[\hat{\mu}_{\textup{occ}}\left|\{\rho(X_{t})\}_{t=1}^{n}\right.\right]\right)
\]

Working with the expectation on the left of the sum:
\[
\mathbb{E}\left[\hat{\mu}_{\textup{occ}}\left|\{\rho(X_{t})\}_{t=1}^{n}\right.\right]=\frac{1}{n}\sum_{t=1}^{n}\mathbb{E}\left[\mathds{1}\{S_{t}=0\}f(X_{t})\left|\{\rho(X_{t})\}_{t=1}^{n}\right.\right]+\mathbb{E}\left[\mathds{1}\{S_{t}=1\}f(Y_{t})\left|\{\rho(X_{t})\}_{t=1}^{n}\right.\right]
\]
Inspecting the first term in the summand:
\begin{align*}
\mathbb{E}\left[\mathds{1}\{S_{t}=0\}f(X_{t})\left|\{\rho(X_{t})\}_{t=1}^{n}\right.\right] & =\mathbb{E}\left[\mathds{1}\{S_{t}=0\}\left|\{\rho(X_{t})\}_{t=1}^{n}\right.\right]\mathbb{E}\left[f(X_{t})\left|\{\rho(X_{t})\}_{t=1}^{n}\right.\right]\\
 & =\left(1-\alpha(\rho(X_{t}))\right)\mu_{\rho(X_{t})}
\end{align*}
where the first line comes from the conditional independence $\{S_{t}\}\bot\{X_{t}\}\left|\{\rho(X_{t})\}\right.$.
Inspecting the second term:
\begin{align*}
\mathbb{E}\left[\mathds{1}\{S_{t}=1\}f(Y_{t})\left|\{\rho(X_{t})\}_{t=1}^{n}\right.\right] & =\mathbb{E}\left[\mathds{1}\{S_{t}=1\}\left|\{\rho(X_{t})\}_{t=1}^{n}\right.\right]\mathbb{E}\left[f(Y_{t})\left|\{\rho(X_{t})\}_{t=1}^{n}\right.\right]\\
 & =\alpha(\rho(X_{t}))\mu_{\rho(X_{t})}
\end{align*}
where the first line comes from the conditional independence $\{S_{t}\}\bot\{Y_{t}\}\left|\{\rho(X_{t})\}\right.$.
Combining the two terms gives
\[
\mathbb{E}\left[\hat{\mu}_{\textup{occ}}\left|\{\rho(X_{t})\}_{t=1}^{n}\right.\right]=\frac{1}{n}\sum_{t=1}^{n}\mu_{\rho(X_{t})}=\frac{1}{n}\sum_{t=1}^{n}\overrightarrow{P}f(X_{t})
\]
and hence 
\[
\textup{Var}\left(\mathbb{E}\left[\hat{\mu}_{\textup{occ}}\left|\{\rho(X_{t})\}_{t=1}^{n}\right.\right]\right)=\textup{Var}\left(\frac{1}{n}\sum_{t=1}^{n}\overrightarrow{P}f(X_{t})\right)
\]
Now we work with the first term in the variance decomposition:
\begin{align*}
\textup{Var}\left(\hat{\mu}_{\textup{occ}}\left|\{\rho(X_{t})\}_{t=1}^{n}\right.\right) & =\frac{1}{n^{2}}\sum_{t=1}^{n}\textup{Var}\left(f_{\textup{occ}}(X_{t},S_{t},Y_{t})\left|\{\rho(X_{t})\}_{t=1}^{n}\right.\right)\\
 & \qquad\qquad+\frac{1}{n^{2}}\sum_{t\neq t'}\textup{Cov}\left(f_{\textup{occ}}(X_{t},S_{t},Y_{t}),f_{\textup{occ}}(X_{t'},S_{t'},Y_{t'})\left|\{\rho(X_{t})\}_{t=1}^{n}\right.\right)
\end{align*}
Inspecting the summand in the first sum:
\begin{align*}
\textup{Var}\left(f_{\textup{occ}}(X_{t},S_{t},Y_{t})\left|\{\rho(X_{t})\}_{t=1}^{n}\right.\right) & =\textup{Var}\left(\mathds{1}\{S_{t}=0\}f(X_{t})+\mathds{1}\{S_{t}=1\}f(Y_{t})\left|\{\rho(X_{t})\}_{t=1}^{n}\right.\right)\\
 & =\mathbb{E}\left[\mathds{1}\{S_{t}=0\}f(X_{t})^{2}+\mathds{1}\{S_{t}=1\}f(Y_{t})^{2}\left|\{\rho(X_{t})\}_{t=1}^{n}\right.\right]\\
 & \qquad-\mathbb{E}\left[\mathds{1}\{S_{t}=0\}f(X_{t})+\mathds{1}\{S_{t}=1\}f(Y_{t})\left|\{\rho(X_{t})\}_{t=1}^{n}\right.\right]^{2}\\
 & =\left(1-\alpha(\rho(X_{t}))\right)\mathbb{E}[f(X_{t})^{2}\left|\{\rho(X_{t})\}_{t=1}^{n}\right.]+\alpha(\rho(X_{t}))\mathbb{E}[f(Y_{t})^{2}\left|\{\rho(X_{t})\}_{t=1}^{n}\right.]\\
 & \qquad-\left((1-\alpha(\rho(X_{t}))\right)\mu_{\rho(X_{t})}+\alpha(\rho(X_{t}))\mu_{\rho(X_{t})})^{2}\\
 & =\left(1-\alpha(\rho(X_{t}))\right)(\sigma_{\rho(X_{t})}^{2}+\mu_{\rho(X_{t})}^{2})+\alpha(\rho(X_{t}))(\sigma_{\rho(X_{t})}^{2}+\mu_{\rho(X_{t})}^{2})-\mu_{\rho(X_{t})}^{2}\\
 & =\sigma_{\rho(X_{t})}^{2}
\end{align*}
where the third equality comes from the conditional independences
$\{S_{t}\}\bot\{X_{t}\}\left|\{\rho(X_{t})\}\right.$ and $\{S_{t}\}\bot\{Y_{t}\}\left|\{\rho(X_{t})\}\right.$.
Inspecting the summand in the second sum:
\begin{align*}
\textup{Cov}\left(f_{\textup{occ}}(X_{t},S_{t},Y_{t}),f_{\textup{occ}}(X_{t'},S_{t'},Y_{t'})\left|\{\rho(X_{t})\}_{t=1}^{n}\right.\right) & =\textup{Cov}\left(\mathds{1}\{S_{t}=0\}f(X_{t}),\mathds{1}\{S_{t'}=0\}f(X_{t'})\left|\{\rho(X_{t})\}_{t=1}^{n}\right.\right)\\
 & =\mathbb{E}\left[\mathds{1}\{S_{t}=0\}f(X_{t})\mathds{1}\{S_{t'}=0\}f(X_{t'})\left|\{\rho(X_{t})\}_{t=1}^{n}\right.\right]\\
 & \qquad-\mathbb{E}\left[\mathds{1}\{S_{t}=0\}f(X_{t})\left|\{\rho(X_{t})\}_{t=1}^{n}\right.\right]\mathbb{E}\left[\mathds{1}\{S_{t'}=0\}f(X_{t'})\left|\{\rho(X_{t})\}_{t=1}^{n}\right.\right]\\
 & =\left(1-\alpha(\rho(X_{t}))\right)\left(1-\alpha(\rho(X_{t'}))\right)\mathbb{E}\left[f(X_{t})f(X_{t'})\left|\{\rho(X_{t})\}_{t=1}^{n}\right.\right]\\
 & \qquad-\left(1-\alpha(\rho(X_{t}))\right)\left(1-\alpha(\rho(X_{t'}))\right)\mu_{\rho(X_{t})}\mu_{\rho(X_{t'})}\\
 & =\left(1-\alpha(\rho(X_{t}))\right)\left(1-\alpha(\rho(X_{t'}))\right)\textup{Cov}\left(f(X_{t}),f(X_{t'})\left|\{\rho(X_{t})\}_{t=1}^{n}\right.\right)\\
 & =\textup{Cov}\left(f_{a}(X_{t}),f_{a}(X_{t'})\left|\{\rho(X_{t})\}_{t=1}^{n}\right.\right)
\end{align*}
where the third equality comes from the conditional independences
$\{S_{t}\}\bot\{X_{t}\}\left|\{\rho(X_{t})\}\right.$ and $S_{i}\bot S_{j}\left|\{\rho(X_{t})\}\right.$
for all $i,j\in[n]$. Using the law of total covariance we have
\begin{align*}
\mathbb{E}\left[\textup{Cov}\left(f_{a}(X_{t}),f_{a}(X_{t'})\left|\{\rho(X_{t})\}_{t=1}^{n}\right.\right)\right] & =\textup{Cov}\left(f_{a}(X_{t}),f_{a}(X_{t'})\right)-\textup{Cov}\left(\mathbb{E}\left[f_{a}(X_{t})\left|\{\rho(X_{t})\}_{t=1}^{n}\right.\right],\mathbb{E}\left[f_{a}(X_{t'})\left|\{\rho(X_{t})\}_{t=1}^{n}\right.\right]\right)\\
 & =\textup{Cov}\left(f_{a}(X_{t}),f_{a}(X_{t'})\right)-\textup{Cov}\left(\overrightarrow{P}f_{a}(X_{t}),\overrightarrow{P}f_{a}(X_{t'})\right)
\end{align*}
Combining everything gives
\[
\mathbb{E}\left[\textup{Var}\left(\hat{\mu}_{\textup{occ}}\left|\{\rho(X_{t})\}_{t=1}^{n}\right.\right)\right]=\frac{1}{n}\mathbb{E}[\sigma_{\rho(X_{t})}^{2}]+\frac{1}{n^{2}}\sum_{t\neq t'}\textup{Cov}\left(f_{a}(X_{t}),f_{a}(X_{t'})\right)-\textup{Cov}\left(\overrightarrow{P}f_{a}(X_{t}),\overrightarrow{P}f_{a}(X_{t'})\right)
\]
We have
that $\mathbb{E}[\sigma_{\rho(X_{t})}^{2}]=\textup{Var}_{P}(\overleftarrow{P}f(X))$
which gives the desired expression, when combined with $\textup{Var}\left(\mathbb{E}\left[\hat{\mu}_{\textup{occ}}\left|\{\rho(X_{t})\}_{t=1}^{n}\right.\right]\right)$ and the results of Proposition \ref{prop:ideal_estimator}.

\subsection{Proofs for Section \ref{section:inherited_theoretical_properties_of_the_occlusion_process}}

\subsubsection{Proof of Proposition \ref{prop:inherited_reversibility}\label{proof:inherited_reversibility}}

For $P_{\textup{occ}}$-reversibility we need that $P_{\textup{occ}}(dx,s,dy)K_{\textup{occ}}((dx,s,dy)\to(dx',s',dy'))=P_{\textup{occ}}(dx',s',dy')K_{\textup{occ}}((dx',s',dy')\to(dx,s,dy))$
for all $dx,dx',dy,dy'\in\mathcal{X}$ and $s,s'\in\{0,1\}$. For
notational simplicity we define
\[
A\left(s\left|x\right.\right):=\alpha(\rho(x))\mathds{1}\{s=1\}+\left(1-\alpha(\rho(x))\right)\mathds{1}\{s=0\}
\]
i.e. the probability mass function of $s$ given $x$. We have
\begin{align*}
P_{\textup{occ}}(dx,s,dy)K_{\textup{occ}}((dx,s,dy)\to(dx',s',dy')) & =P(dx)A\left(s\left|x\right.\right)P_{\rho(x)}(dy)K(x\to dx')A\left(s'\left|x'\right.\right)P_{\rho(x')}(dy')\\
 & =P(dx)K(x\to dx')A\left(s'\left|x'\right.\right)P_{\rho(x')}(dy')A\left(s\left|x\right.\right)P_{\rho(x)}(dy)\\
 & =P(dx')K(x'\to dx)A\left(s'\left|x'\right.\right)P_{\rho(x')}(dy')A\left(s\left|x\right.\right)P_{\rho(x)}(dy)\\
 & =P(dx')A\left(s'\left|x'\right.\right)P_{\rho(x')}(dy')K(x'\to dx)A\left(s\left|x\right.\right)P_{\rho(x)}(dy)\\
 & =P_{\textup{occ}}(dx',s',dy')K_{\textup{occ}}((dx',s',dy')\to(dx,s,dy))
\end{align*}
where the third equality comes from the $P$-reversibility of $K$.

\subsubsection{Proof of Proposition \ref{prop:inherited_l2}\label{proof:inherited_l2}}

We directly observe
\begin{align*}
\int_{\mathsf{X}}\sum_{s=0}^{1}\int_{\mathsf{X}}f_{\textup{occ}}(x,s,y)^{2}P_{\textup{occ}}(dx,s,dy) & =\int_{\mathsf{X}}\int_{\mathsf{X}}f(y)^{2}\left(1-\alpha(\rho(x))\right)P(dx)P_{\rho(x)}(dy)+\int_{\mathsf{X}}\int_{\mathsf{X}}f(x)^{2}\alpha(\rho(x))P(dx)P_{\rho(x)}(dy)\\
 & =\sum_{i=1}^{R}P(\mathsf{X}_{i})\left(1-\alpha(i)\right)P_{i}(f^{2})+P(\mathsf{X}_{i})\alpha(i)P_{i}(f^{2})\\
 & =P(f^{2})<\infty
\end{align*}
where the last line is due to the fact that $f\in L^{2}(P)$.

\subsubsection{Proof of theorem \ref{thm:LLN_inheritance}}\label{proof:LLN_inheritance}

\paragraph{1. $\Rightarrow$ 2.}Let $K$ be the kernel of the Markov chain $\{X_{t}\}_{t=1}^{n}$.
\cite[Proposition 3.5]{douc:2023} says the LLN stated in statement 1. is
equivalent to the fact that for all functions $h:\mathsf{X}\to\mathbb{R}$
if $Kh\equiv h$ then $h$ is constant. Let $h_{\textup{occ}}:\mathsf{X}\times\{0,1\}\times\mathsf{X}\to\mathbb{R}$
be such that $K_{\textup{occ}}h_{\textup{occ}}\equiv h_{\textup{occ}}$.
Then for all $(x,s,y)\in\mathsf{X}\times\{0,1\}\times\mathsf{X}$
\[
\int_{\mathsf{X}}\sum_{s'=0}^{1}\int_{\mathsf{X}}h_{\textup{occ}}(x',s',y')K_{\textup{occ}}((x,s,y)\to(dx',s',dy'))=h_{\textup{occ}}(x,s,y)
\]
The fact that $K_{\textup{occ}}((x,s,y)\to(dx',s',dy'))$ is independent
of $s$ and $y$ means that $h_{\textup{occ}}$ is a function of $x$
only. Therefore the above equation is equivalent to 
\begin{align*}
\int_{\mathsf{X}}\sum_{s'=0}^{1}\int_{\mathsf{X}}h_{\textup{occ}}(x')K_{\textup{occ}}((x,s,y)\to(dx',s',dy')) & =h_{\textup{occ}}(x)\\
\Rightarrow Kh_{\textup{occ}}(x) & =h_{\textup{occ}}(x)
\end{align*}
and so $h_{\textup{occ}}$ is constant by hypothesis. Therefore \cite[Proposition 3.5]{douc:2023}  asserts an LLN for the occlusion process.

\paragraph{2. $\Rightarrow$ 1.} Assuming statement 2. means that $K_\textup{occ} h_\textup{occ} \equiv h_\textup{occ}$ implies $h_\textup{occ}$ is a constant. Here we use the fact that applying $K_\textup{occ}$ to a function of only its first variable is the same as applying $K$ to it. Thus we have that for any $h:\mathsf{X} \to \mathbb{R}$
\[
\begin{split}
    Kh \equiv h &\Rightarrow K_\textup{occ} h \equiv h\\
    &\Rightarrow h\textup{ is constant}
\end{split}
\]
and we are done.

\subsubsection{Proof of theorem \ref{thm:inherited_normed_function_convergence}\label{proof:inherited_normed_function_convergence}}

Let $(\mathsf{G},\|.\|_{\mathsf{G}})\in\mathcal{C}_{\|.\|}$ and let
$g(x,s,y)=\mathds{1}\{s=0\}f(x)+\mathds{1}\{s=1\}f(y)\in(\mathsf{G},\|.\|_{\mathsf{G}})$
as in the theorem statement. First note that $P_{\textup{occ}}(g)=P(f_{\alpha})$.
Then we have that
\begin{align*}
K_{\textup{occ}}g(x,s,y) & =\int_{\mathsf{X}}\sum_{s'=0}^{1}\int_{\mathsf{X}}(\mathds{1}\{s'=0\}f(x')+\mathds{1}\{s'=1\}f(y'))K_{\textup{occ}}((x,s,y)\to(dx',s',dy'))\\
 & =\int_{\mathsf{X}}\int_{\mathsf{X}}\left(1-\alpha(\rho(x'))\right)f(x')K(x\to dx')P_{\rho(x')}(dy')\\
 & \qquad\qquad+\int_{\mathsf{X}}\int_{\mathsf{X}}\alpha(\rho(x'))f(y')K(x\to dx')P_{\rho(x')}(dy')\\
 & =\int_{\mathsf{X}}\left(1-\alpha(\rho(x'))\right)f(x')K(x\to dx')+\int_{\mathsf{X}}\alpha(\rho(x'))P_{\rho(x')}(f)K(x\to dx')\\
 & =Kf_{\alpha}(x)
\end{align*}
Since $K_{\textup{occ}}=K$ when acting on functions who only depend
on $x$ we have that $K_{\textup{occ}}^{t}g=K^{t}f_{\alpha}$ for
$t\in\mathbb{N}\backslash\{0\}$. Therefore
\begin{align*}
\|K_{\textup{occ}}^{t}g-P_{\textup{occ}}(g)\|_{\mathsf{G}} & =\|K^{t}f_{\alpha}-P(f_{\alpha})\|\\
 & \leq C_{f_{\alpha}}r(t)
\end{align*}
where the first line is due to the fact that $\|g\|_{\mathsf{G}}=\|g\|$
for all functions $g$ solely of their first argument and the second
line is by hypothesis.

\subsubsection{Proof that $C_{f_{\alpha}}\protect\leq C_{f}$ in \ref{ex:sup_norm}\label{proof:sup_norm}}

Note that $C_{f}:=C\|f-P(f)\|$ with $C>0$ where $\|.\|$ is the
sup norm. Define $f_{0}:=f-P(f)$. Then
\begin{align*}
C_{f_{\alpha}} & =C\|f_{\alpha}-P(f_{\alpha})\|\\
 & =C\sup_{x\in\mathsf{X}}\left|\left(1-\alpha(\rho(x))\right)f(x)+\alpha(\rho(x))\overrightarrow{P}f(x)-P(f)\right|\\
 & =C\sup_{x\in\mathsf{X}}\left|\left(1-\alpha(\rho(x))\right)f_{0}(x)+\alpha(\rho(x))\overrightarrow{P}f_{0}(x)\right|\\
 & \leq C\sup_{x\in\mathsf{X}}\left(1-\alpha(\rho(x))\right)\left|f_{0}(x)\right|+\alpha(\rho(x))\left|\overrightarrow{P}f_{0}(x)\right|\\
 & \leq C\|f_{0}\|
\end{align*}
where the second equality comes from the fact that $P(f_{\alpha})=P(f)$,
the third equality comes from the fact that $P(\overrightarrow{P}f)=P(f)$,
the first inequality comes from Jensen's inequality, and the final
inequality comes from the fact that $\left|f_{0}(x)\right|$ and $\left|\overrightarrow{P}f_{0}(x)\right|$
are upper bounded by $\|f_{0}\|$.

\subsubsection{Proof of theorem \ref{thm:inherited_IPM_convergence}\label{proof:inherited_IPM_convergence}}

Say that $\nu$ is the distribution of the first state $(X_{1},S_{1},Y_{1})$
of the occlusion process, such that $\mu$ is the marginal of its
first component. Then the result follows from the fact that $\nu K_{\textup{occ}}^{t}(g)=\mu K^{t}(f_{\alpha})$
and $P_{\textup{occ}}(g)=P(f_{\alpha})$ for all $g\in\mathsf{G}$
and $\mathsf{G}\in\mathcal{C}$. These results are derived in {[}section
6.3.1{]}. In particular, we have that
\begin{align*}
D_{\mathsf{G}}(\nu K_{\textup{occ}}^{t},P_{\textup{occ}}) & =\sup_{g\in\mathsf{G}}\left|\nu K_{\textup{occ}}^{t}(g)-P_{\textup{occ}}(g)\right|\\
 & =\sup_{f\in\mathsf{F}}\left|\mu K^{t}(f_{\alpha})-P(f_{\alpha})\right|\\
 & \leq C_{\mu}r(t)
\end{align*}
where the second line comes from the fact that as $f$ ranges over
$\mathsf{F}$, $g$ ranges over a subset of $\mathsf{G}$. The final
line is by hypothesis.

\subsubsection{Proof of theorem \ref{thm:inherited_geometric_ergodicity}\label{proof:inherited_geometric_ergodicity}}

\cite[Theorem 1 viii)]{gallegos-herrada:2024}
has that the geometric ergodicity of a Markov kernel $K$ is equivalent
to the existence of a $P$-almost everywhere measurable function $V:\mathsf{X}\to[1,\infty]$,
a small set $S\in\mathcal{X}$ and constants $\lambda<1$ and $b<\infty$
with
\[
KV(x)\leq\lambda V(x)+b\mathds{1}\{x\in S\}
\]

Defining $V_{\textup{occ}}(x,s,y):=V(x)$ and $S_{\textup{occ}}:=S\times\{0,1\}\times\mathsf{X}$
we prove that $V_{\textup{occ}}$ and $S_{\textup{occ}}$ satisfy
an inequality such as the above with $K_{\textup{occ}}$. Firstly
\begin{align*}
K_{\textup{occ}}V_{\textup{occ}}(x,s,y) & =\int_{\mathsf{X}}\sum_{s'=0}^{1}\int_{\mathsf{X}}K_{\textup{occ}}((x,s,y)\to(dx',s',dy'))V_{\textup{occ}}(x',s',y')\\
 & =\int_{\mathsf{X}}\sum_{s'=0}^{1}\int_{\mathsf{X}}K(x\to dx')\left(\left(1-\alpha(\rho(x'))\right)\mathds{1}\{s'=0\}+\alpha(\rho(x'))\mathds{1}\{s'=1\}\right)P_{\rho(x')}(dy')V(x)\\
 & =\int_{\mathsf{X}}\int_{\mathsf{X}}K(x\to dx')V(x)P_{\rho(x')}(dy')\\
 & =\int_{\mathsf{X}}K(x\to dx')V(x)=KV(x)
\end{align*}
Hence $K_{\textup{occ}}V_{\textup{occ}}(x,s,y)\leq\lambda V(x)+b\mathds{1}\{x\in S\}$.
Noting that $V(x)\equiv V_{\textup{occ}}(x,s,y)$ and $\mathds{1}\{x\in S\}\equiv\mathds{1}\{(x,s,y)\in S_{\textup{occ}}\}$
along with the fact that $V_{\textup{occ}}$ is $P_{\textup{occ}}$-almost
everywhere measurable gives the result.

\subsubsection{Proof of corollary \ref{cor:CLT_from_geometric_ergodicity}\label{proof:CLT_from_geometric_ergodicity}}

Since $K$ is geometrically ergodic and $P$-reversible Proposition
\ref{prop:inherited_reversibility} and Theorem \ref{thm:inherited_geometric_ergodicity}
imply that $K_{\textup{occ}}$ is geometrically ergodic and $P_{\textup{occ}}$-reversible.
Similarly Proposition \ref{prop:inherited_l2} has that $f_{\textup{occ}}\in L^{2}(P_{\textup{occ}})$.
Therefore \cite[Corollary 2.1]{roberts:97} implies
that $\hat{\mu}_{\textup{occ}}$ admits the CLT as in the statement
of the corollary.

For the finiteness of the asymptotic variance, without loss of generality
we work with $f_{0}:=f_{\textup{occ}}-\mu$ since 
\[
\lim_{n\to\infty}n\textup{Var}(n^{-1}\sum_{t=1}^{n}f_{\textup{occ}}(X_{t},S_{t},Y_{t}))=\lim_{n\to\infty}n\textup{Var}(n^{-1}\sum_{t=1}^{n}f_{0}(X_{t},S_{t},Y_{t}))
\]
Expanding the variance then gives
\[
\lim_{n\to\infty}n\textup{Var}(n^{-1}\sum_{t=1}^{n}f_{0}(X_{t},S_{t},Y_{t}))=\|f_{0}\|_{P_{\textup{occ}}}^{2}+2\lim_{n\to\infty}\sum_{k=1}^{n-1}\frac{n-k}{n}\langle f_{0},K^{k}f_{0}\rangle_{P_{\textup{occ}}}
\]
The sum in the limit is a Ces\`aro sum and so converges to $\sum_{k=1}^{\infty}\langle f_{0},K^{k}f_{0}\rangle_{P_{\textup{occ}}}$
if $\sum_{k=1}^{\infty}|\langle f_{0},K^{k}f_{0}\rangle_{P_{\textup{occ}}}|<\infty$.
Geometric ergodicity along with reversibility implies a spectral gap
\cite[Theorem 1 xxx)]{gallegos-herrada:2024} which
means that $|\langle f_{0},K^{k}f_{0}\rangle_{P_{\textup{occ}}}|\leq\lambda^{k}\|f_{0}\|_{P_{\textup{occ}}}^{2}$
for some $\lambda\in[0,1)$. This gives absolute convergence and hence
the convergence of the Ces\`aro sum. Noting that $\|f_{0}\|_{P_{\textup{occ}}}^{2}=\textup{Var}_{P_{\textup{occ}}}(f)$
and $\langle f_{0},K^{k}f_{0}\rangle_{P_{\textup{occ}}}=\textup{Cov}_{P_{\textup{occ}}}(f_{\textup{occ}},K^{k}f_{\textup{occ}})$
completes the proof.

\subsubsection{Proof of Proposition \ref{prop:asymptotic_occlusion_variance_equivalence}}\label{proof:asymptotic_occlusion_variance_equivalence}

From Proposition \ref{prop:variance_mu_occ} we have that
\begin{equation}\label{eqn:asy_variance_mu_hat_occ_decomposition}
    \lim_{n\to\infty}n\textup{Var}(\hat{\mu}_\textup{occ}) = \lim_{n\to\infty}n\textup{Var}(\hat{\mu}_\textup{ideal}) + 2 \lim_{n\to\infty} \sum_{k = 1}^{n - 1}\frac{n - k}{n}C_k
\end{equation}
where
\[
C_k := \textup{Cov}_P(f_a(X), K^kf_a(X)) - \textup{Cov}_P(\overrightarrow{P}f_a(X), K^k\overrightarrow{P}f_a(X))
\]
and $f_a(x):= (1 - \alpha(\rho(x)))f(x)$, so long as the two limits on the right hand side of (\ref{eqn:asy_variance_mu_hat_occ_decomposition}) converge: we will see that they do.

For the first limit, Proposition \ref{prop:ideal_estimator} dictates that
\[
\lim_{n\to\infty}n\textup{Var}(\hat{\mu}_\textup{ideal}) = \textup{Var}_P(\overleftarrow{P}f) + \lim_{n \to\infty}\textup{Var}\left(\frac{1}{n}\sum_{t=1}^n\overrightarrow{P}f(X_t)\right)
\]
Geometric ergodicity and reversibility implies a spectral gap \cite[Theorem 1 xxx)]{gallegos-herrada:2024}, which itself implies that the above limit on the right hand side of the equation can be expressed as
\[
\lim_{n \to\infty}\textup{Var}\left(\frac{1}{n}\sum_{t=1}^n\overrightarrow{P}f(X_t)\right) = \textup{Var}_P(\overrightarrow{P}f) + 2\sum_{k = 1}^\infty \textup{Cov}_P(\overrightarrow{P}f(X), K^k\overrightarrow{P}f(X))
\]

Now to address the second limit on the right hand side of the equation (\ref{eqn:asy_variance_mu_hat_occ_decomposition}). The expression inside the limit is a Ces\`aro sum, and hence converges when $\sum_{k=1}^\infty|C_k|<\infty$. The existence of a spectral gap for $K$ implies the existence of a $\lambda \in [0, 1)$ such that
\begin{equation*}
    \begin{split}
        |C_k| &= |\langle f_a + \overrightarrow{P}f_a, K^k(f_a - \overrightarrow{P}f_a)\rangle|\\
        &\leq \lambda^k|\langle f_a + \overrightarrow{P}f_a, f_a - \overrightarrow{P}f_a\rangle|
    \end{split}
\end{equation*}
and hence the sum converges absolutely.

\subsection{Proofs for Section 4}

\subsubsection{Proof that $Y$ is sampled from $P$ restricted to $\mathsf{X}_C$}\label{proof:Y_from_P_restricted}

We have that 
\begin{align*}
\mathbb{P}\left(Y\in A\left|U\leq\frac{1}{C}\frac{d\tilde{P}}{d\tilde{Q}}(Y)\cap Y\in\mathsf{X}_{C}\right.\right) & =\frac{\mathbb{E}_{Y,U}\left[\mathds{1}\{Y\in A\}\mathds{1}\{U\leq\frac{1}{C}\frac{d\tilde{P}}{d\tilde{Q}}(Y)\}\mathds{1}\{Y\in\mathsf{X}_{C}\}\right]}{\mathbb{E}_{Y,U}\left[\mathds{1}\{U\leq\frac{1}{C}\frac{d\tilde{P}}{d\tilde{Q}}(Y)\}\mathds{1}\{Y\in\mathsf{X}_{C}\}\right]}\\
 & =\frac{\mathbb{E}_{Y}\left[\mathds{1}\{Y\in A\}\mathbb{E}_{U}\left[\mathds{1}\{U\leq\frac{1}{C}\frac{d\tilde{P}}{d\tilde{Q}}(Y)\}\right]\mathds{1}\{Y\in\mathsf{X}_{C}\}\right]}{\mathbb{E}_{Y}\left[\mathbb{E}_{U}\left[\mathds{1}\{U\leq\frac{1}{C}\frac{d\tilde{P}}{d\tilde{Q}}(Y)\}\right]\mathds{1}\{Y\in\mathsf{X}_{C}\}\right]}\\
 & =\frac{\mathbb{E}_{Y}\left[\mathds{1}\{Y\in A\}\frac{1}{C}\frac{d\tilde{P}}{d\tilde{Q}}(Y)\mathds{1}\{Y\in\mathsf{X}_{C}\}\right]}{\mathbb{E}_{Y}\left[\frac{1}{C}\frac{d\tilde{P}}{d\tilde{Q}}(Y)\mathds{1}\{Y\in\mathsf{X}_{C}\}\right]}\\
 & =\frac{\mathbb{E}_{Y\sim P}\left[\mathds{1}\{Y\in A\}\mathds{1}\{Y\in\mathsf{X}_{C}\}\right]}{\mathbb{E}_{Y\sim P}\left[\mathds{1}\{Y\in\mathsf{X}_{C}\}\right]}=\mathbb{P}_{Y\sim P}(Y\in A\left|Y\in\mathsf{X}_{C}\right.)
\end{align*}

\subsection{Proofs for Appendix B}

\subsubsection{Proof of theorem \ref{thm:inherited_normed_measure_convergence}\label{proof:inherited_normed_measure_convergence}}

First note that for all $B=B_{x}\times B_{s}\times B_{y}\in\mathcal{X}\times2^{\{0,1\}}\times\mathcal{X}$
we have
\begin{align*}
\nu K_{\textup{occ}}(B) & =\int_{\mathsf{X}}\sum_{s=0}^{1}\int_{\mathsf{X}}\nu(dx,s,dy)K_{\textup{occ}}((x,s,y)\to B)\\
 & =\int_{\mathsf{X}}\sum_{s=0}^{1}\int_{\mathsf{X}}\mu(dx)q(s,dy\left|x\right.)K_{\textup{occ}}((x,s,y)\to B)\\
 & =\int_{x\in\mathsf{X}}\mu(dx)K_{\textup{occ}}((x,s,y)\to B)\\
 & =\int_{x\in\mathsf{X}}\mu(dx)\int_{x'\in B_{x}}K(x\to dx')\left(\alpha(\rho(x'))\mathds{1}\{1\in B_{s}\}+\left(1-\alpha(\rho(x'))\right)\mathds{1}\{0\in B_{s}\}\right)P_{\rho(x')}(B_{y})
\end{align*}
for $\mu\in(\mathsf{M},\|.\|_{*,\mathsf{M}})$ where the second equality
is by hypothesis and the third equality comes from the fact that $K_{\textup{occ}}((x,s,y)\to B)$
is independent of $s$ and $y$. Continuing, we have
\begin{align*}
\nu K_{\textup{occ}}(B) & =\sum_{i=1}^{R}\left(\alpha(i)\mathds{1}\{1\in B_{s}\}+\left(1-\alpha(i)\right)\mathds{1}\{0\in B_{s}\}\right)P_{i}(B_{y})\int_{x\in\mathsf{X}}\mu(dx)K(x\to B_{x}\cap\mathsf{X}_{i})\\
 & =\sum_{i=1}^{R}\left(\alpha(i)\mathds{1}\{1\in B_{s}\}+\left(1-\alpha(i)\right)\mathds{1}\{0\in B_{s}\}\right)P_{i}(B_{y})\mu K(B_{x}\cap\mathsf{X}_{i})
\end{align*}
By a similar argument we have that
\[
\nu K_{\textup{occ}}^{t}(B)=\sum_{i=1}^{R}\left(\alpha(i)\mathds{1}\{1\in B_{s}\}+\left(1-\alpha(i)\right)\mathds{1}\{0\in B_{s}\}\right)P_{i}(B_{y})\mu K^{t}(B_{x}\cap\mathsf{X}_{i})
\]
for all $t\in\mathbb{N}\backslash\{0\}$.

Dual norms are defined as suprema of a dual object as evaluated across
the primal space. Therefore we need to inspect how $\nu K_{\textup{occ}}^{t}$
acts on a function $g$ in the normed function space $(\mathsf{G},\|.\|_{\mathsf{G}})\in\mathcal{C}_{\|.\|}$.
\begin{align*}
\nu K_{\textup{occ}}^{t}(g) & =\sum_{i=1}^{R}\int_{\mathsf{X}}\sum_{s=0}^{1}\int_{\mathsf{X}}g(x,s,y)\left(\alpha(i)\mathds{1}\{s=1\}+\left(1-\alpha(i)\right)\mathds{1}\{s=0\}\right)P_{i}(dy)\mu K^{t}(dx)\mathds{1}\{x\in\mathsf{X}_{i}\}\\
 & =\sum_{i=1}^{R}\int_{\mathsf{X}}\int_{\mathsf{X}}f(y)\alpha(i)P_{i}(dy)\mu K^{t}(dx)\mathds{1}\{x\in\mathsf{X}_{i}\}+\int_{\mathsf{X}}\int_{\mathsf{X}}f(x)\left(1-\alpha(i)\right)P_{i}(dy)\mu K^{t}(dx)\mathds{1}\{x\in\mathsf{X}_{i}\}\\
 & =\sum_{i=1}^{R}\alpha(i)\mu K^{t}(P_{i}(f)\mathds{1}\{.\in\mathsf{X}_{i}\})+\left(1-\alpha(i)\right)\mu K^{t}(f\mathds{1}\{.\in\mathsf{X}_{i}\})\\
 & =\mu K^{t}\left(\sum_{i=1}^{R}\alpha(i)P_{i}(f)\mathds{1}\{.\in\mathsf{X}_{i}\}+\left(1-\alpha(i)\right)f\mathds{1}\{.\in\mathsf{X}_{i}\}\right)\\
 & =\mu K^{t}(f_{\alpha})
\end{align*}
Equally, we have that $P_{\textup{occ}}(g)=P(f_{\alpha})$. Therefore
the dual norm has the following form:
\begin{align*}
\|\nu K_{\textup{occ}}^{t}-P_{\textup{occ}}\|_{*,\mathsf{N}} & =\sup_{g\in(\mathsf{G},\|.\|_{\mathsf{G}})}\frac{\left|\left(\nu K_{\textup{occ}}^{t}-P_{\textup{occ}}\right)(g)\right|}{\|g\|_{\mathsf{G}}}\\
 & =\sup_{f\in(\mathsf{F},\|.\|)}\frac{\left|\left(\mu K^{t}-P\right)(f_{\alpha})\right|}{\|g\|_{\mathsf{G}}}
\end{align*}
Where the second equality comes from the fact that as $f$ ranges
over $(\mathsf{F},\|.\|)$, $g$ ranges over $(\mathsf{G},\|.\|_{\mathsf{G}})$.
To place an upper bound on the numerator within the supremum, we have
that $\left|\left(\mu K^{t}-P\right)(f_{\alpha})\right|=\|\mu K^{t}-P\|_{\mathsf{M},*}d(f_{\alpha},\ker(\mu K^{t}-P))$
where $d(f,S):=\inf_{h\in S}\|f-h\|$ for all $f\in(\mathsf{F},\|.\|)$
and $S\subseteq\mathsf{F}$ \cite[Lemma 1.1]{hashimoto:1986}. Since $0\in\ker(\mu K^{t}-P)$ we have that $\left|\left(\mu K^{t}-P\right)(f_{\alpha})\right|\leq\|\mu K^{t}-P\|_{\mathsf{M},*}\|f_{\alpha}\|$
and hence
\begin{align*}
\|\nu K_{\textup{occ}}^{t}-P_{\textup{occ}}\|_{*,\mathsf{N}} & \leq\left(\sup_{f\in(\mathsf{F},\|.\|)}\frac{\|f_{\alpha}\|}{\|g\|_{\mathsf{G}}}\right)\|\mu K^{t}-P\|_{\mathsf{M},*}\\
 & \leq\left(\sup_{f\in(\mathsf{F},\|.\|)}\frac{\|f_{\alpha}\|}{\|g\|_{\mathsf{G}}}\right)C_{\mu}r(t)
\end{align*}
where the final inequality is by hypothesis.

\section{Appendix B: Inherited convergence of the occlusion process in a generic
normed measure space}\label{appendix:normed_measure_convergence}

As stated in Section \ref{subsubsection:convergence_in_a_normed_measure_space} the inherited convergence of the
occlusion process in IPMs was just an example of a more general phenomenon.
Here we detail a result which applies to a more general class of normed
measure spaces. Normed measure spaces are often dual to normed function
spaces. Therefore the following inheritance is stated relative to
two normed function spaces: $(\mathsf{F},\|.\|)$ and $(\mathsf{G},\|.\|_{\mathsf{G}})\in\mathcal{C}_{\|.\|}$
where $\mathcal{C}_{\|.\|}$ is defined as in Section \ref{subsubsection:convergence_in_a_normed_function_space}.
\begin{thm}
\label{thm:inherited_normed_measure_convergence}Say the Markov chain
$\{X_{t}\}_{t=1}^{n}$ converges to $P$ in the normed measure space
$(\mathsf{M},\|.\|_{*,\mathsf{M}})$ dual to the normed function space
$(\mathsf{F},\|.\|)$ with rate function $r(t)$ and constant $C_{\mu}$.
Consider all normed measure spaces $(\mathsf{N},\|.\|_{*,\mathsf{N}})$
dual to the normed function spaces $(\mathsf{G},\|.\|_{\mathsf{G}})\in\mathcal{C}_{\|.\|}$.
Then for all measures $\nu\in(\mathsf{N},\|.\|_{*,\mathsf{N}})$ such
that $\nu(dx,s,dy)=\mu(dx)q(s,dy\left|x\right.)$ with $\mu\in(\mathsf{M},\|.\|_{*,\mathsf{M}})$
for all $(dx,s,dy)\in\mathcal{X}\times\{0,1\}\times\mathcal{X}$ we
have that
\[
\|\nu K_{\textup{occ}}^{t}-P_{\textup{occ}}\|_{*,\mathsf{N}}\leq\left(\sup_{f\in(\mathsf{F},\|.\|)}\frac{\|f_{\alpha}\|}{\|g\|_{\mathsf{G}}}\right)C_{\mu}r(t)
\]
where
\[
f_{\alpha}(x):=\left(1-\alpha(\rho(x))\right)f(x)+\alpha(\rho(x))\overrightarrow{P}f(x)
\]
and
\[
g(x,s,y):=\mathds{1}\{s=0\}f(x)+\mathds{1}\{s=1\}f(y)
\]
\end{thm}

\begin{ex}
Say $\mathsf{X}$ is Hausdorff and compact, and $(\mathsf{F},\|.\|)$
is the space of continuous functions on $\mathsf{X}$ with the sup
norm. Then $(\mathsf{M},\|.\|_{*,\mathsf{M}})$ is the space of regular
countably additive measures with $\|\mu\|_{*,\mathsf{M}}:=\sup_{f\in(\mathsf{F},\|.\|)}\|f\|^{-1}|\mu(f)|$.
So for a given $(\mathsf{G},\|.\|_{\mathsf{G}})\in\mathcal{C}_{\|.\|}$
we have that $g\in(\mathsf{G},\|.\|_{\mathsf{G}})$ when there exists
an $f\in(\mathsf{F},\|.\|)$ such that $g(x,s,y)=\mathds{1}\{s=0\}f(x)+\mathds{1}\{s=1\}f(y)$.
In this case
\begin{align*}
\|g\|_{\mathsf{G}} & =\sup_{(x,s,y)\in\mathsf{X}\times\{0,1\}\times\mathsf{X}}|g(x,s,y)|\\
 & =\sup_{(x,s,y)\in\mathsf{X}\times\{0,1\}\times\mathsf{X}}|\mathds{1}\{s=0\}f(x)+\mathds{1}\{s=1\}f(y)|\\
 & =\max\{\sup_{x\in\mathsf{X}}|f(x)|,\sup_{y\in\mathsf{X}}|f(y)|\}\\
 & =\sup_{x\in\mathsf{X}}|f(x)|=\|f\|
\end{align*}
where in the third equality we have split into the cases $s=0$ and
$s=1$. To work out the new constant for the convergence in $(\mathsf{N},\|.\|_{*,\mathsf{N}})$
of the occlusion process $\{(X_{t},S_{t},Y_{t})\}_{t=1}^{n}$ we have
\begin{align*}
\sup_{f\in(\mathsf{F},\|.\|)}\frac{\|f_{\alpha}\|}{\|g\|_{\mathsf{G}}} & =\sup_{f\in(\mathsf{F},\|.\|)}\frac{\|(1-\alpha(\rho(.)))f+\alpha(\rho(.))\overrightarrow{P}f\|}{\|f\|}\\
 & \leq\sup_{f\in(\mathsf{F},\|.\|)}\frac{\|(1-\alpha(\rho(.)))f\|+\|\alpha(\rho(.))\overrightarrow{P}f\|}{\|f\|}\\
 & \leq\sup_{f\in(\mathsf{F},\|.\|)}\frac{\|f\|+\|\overrightarrow{P}f\|}{\|f\|}\leq2
\end{align*}
where the final inequality is due to the fact that $\|\overrightarrow{P}f\|\leq\|f\|$.
\end{ex}

\section{Appendix C: Additional results for the Ising experiment in Section \ref{subsection:Ising_model}}\label{appendix:additional_results_for_the_Ising_experiment_in_section}

\begin{figure}
    \centering
    \includegraphics[scale = 0.9]{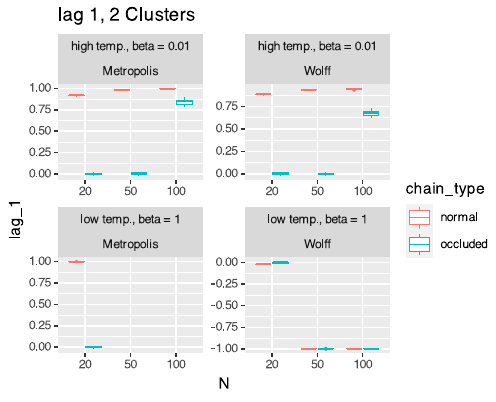}
    \includegraphics[scale = 0.9]{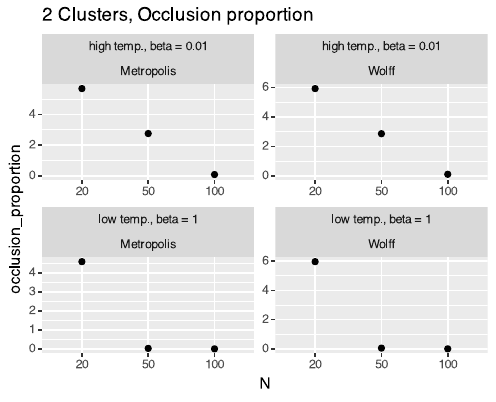}
    \includegraphics[scale = 1.3]{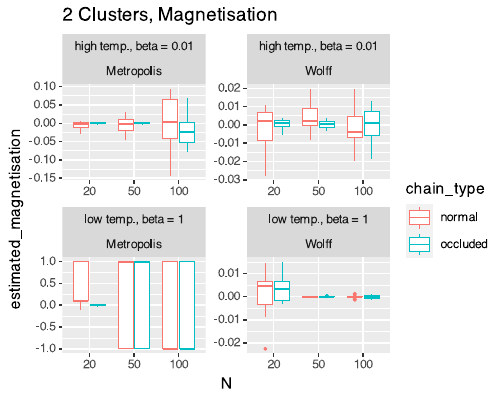}
    \caption{Three graphs comparing the performance of the occlusion process with the Metropolis and Wolff algorithms on the Ising model at a variety of temperatures, for a variety of graph sizes, each generated using the stochastic block model with 2 communities. In every case the horizontal axes show the number of vertices $N$ in the graphs. Bottom: the vertical axes denote the algorithm's estimates of the expected magnetisation. Top left: the vertical axes denote the lag 1 autocorrelation coefficient of the magnetisation over the states produced by the algorithms. Top right: the vertical axes show the number of samples from the $P_i$'s in Algorithm \ref{alg:embarrassingly_parallel} divided by the number of states in the Markov chain $n$. We magnify the estimated magnetisation plot for ease of comprehension, the top two plots then help to explain the phenomena in the bottom plot.}
    \label{fig:Ising_results_2_clusters}
\end{figure}

\begin{figure}
    \centering
    \includegraphics[scale = 0.9]{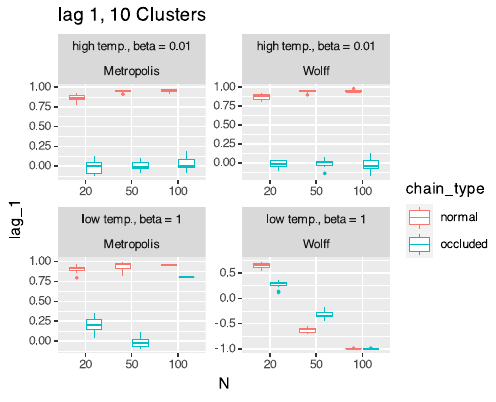}
    \includegraphics[scale = 0.9]{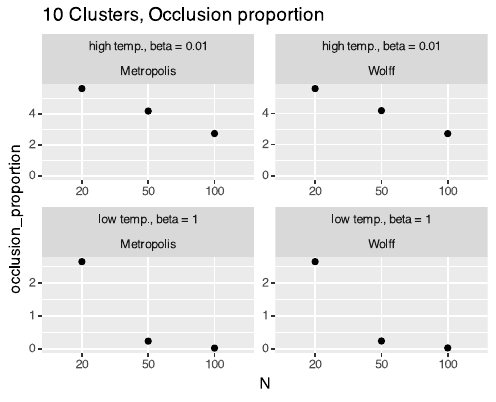}
    \includegraphics[scale = 1.3]{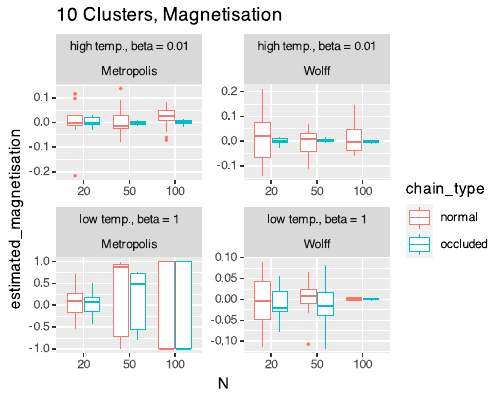}
    \caption{Three graphs comparing the performance of the occlusion process with the Metropolis and Wolff algorithms on the Ising model at a variety of temperatures, for a variety of graph sizes, each generated using the stochastic block model with 10 communities. In every case the horizontal axes show the number of vertices $N$ in the graphs. Bottom: the vertical axes denote the algorithm's estimates of the expected magnetisation. Top left: the vertical axes denote the lag 1 autocorrelation coefficient of the magnetisation over the states produced by the algorithms. Top right: the vertical axes show the number of samples from the $P_i$'s in Algorithm \ref{alg:embarrassingly_parallel} divided by the number of states in the Markov chain $n$. We magnify the estimated magnetisation plot for ease of comprehension, the top two plots then help to explain the phenomena in the bottom plot.}
    \label{fig:Ising_results_10_clusters}
\end{figure}

\bibliographystyle{unsrtnat}
\bibliography{references}  






\end{document}